\documentclass[aps, prl, superscriptaddress, floatfix, longbibliography, preprint]{revtex4-2}

\usepackage{amsfonts, amssymb, amsmath}
\usepackage{graphicx}
\usepackage{float}
\usepackage[plain]{fancyref}
\usepackage{placeins}
\usepackage{hyperref}
\usepackage{color,soul}
\usepackage{siunitx}
\usepackage{natbib}
\usepackage[version=3]{mhchem}
\usepackage{physics}
\usepackage{pdfrender}
\usepackage{mathtools}
\usepackage{cases}
\usepackage{verbatim}
\definecolor{myorange}{cmyk}{0.0149, 0.7523, 0.7816, 0.003}
\definecolor{mypurple}{cmyk}{0.6638, 0.9652, 0, 0}
\usepackage[normalem]{ulem}

\DeclareSIUnit\angstrom{\text {Å}}
\newcommand{\VQD}{\ensuremath{V_\mathrm{QD}}}
\newcommand{\VHi}{\ensuremath{V_\mathrm{H1}}}
\newcommand{\VHii}{\ensuremath{V_\mathrm{H2}}}
\newcommand{\VL}{\ensuremath{V_\mathrm{L}}}
\newcommand{\VR}{\ensuremath{V_\mathrm{R}}}
\newcommand{\IL}{\ensuremath{I_\mathrm{L}}}
\newcommand{\IR}{\ensuremath{I_\mathrm{R}}}
\newcommand{\gL}{\ensuremath{g_\mathrm{L}}}
\newcommand{\gR}{\ensuremath{g_\mathrm{R}}}
\newcommand{\Vbias}{\ensuremath{V_\mathrm{bias}}}
\newcommand{\Imeas}{\ensuremath{I_\mathrm{meas}}}
\newcommand{\Ibias}{\ensuremath{I_\mathrm{bias}}}
\newcommand{\Vmeas}{\ensuremath{V_\mathrm{meas}}}
\newcommand{\Isw}{\ensuremath{I_\mathrm{sw}}}
\newcommand{\Ic}{\ensuremath{I_\mathrm{c}}}

\newcommand{\NL}{\ensuremath{\mathrm{N}_\textrm{L}}}
\newcommand{\NR}{\ensuremath{\mathrm{N}_\textrm{R}}}
\newcommand{\SL}{\ensuremath{\mathrm{S}_\textrm{1}}}
\newcommand{\SR}{\ensuremath{\mathrm{S}_\textrm{2}}}

\begin{document}

\title{Supercurrent through an Andreev trimer}

\author{Alberto~Bordin}
\thanks{These authors contributed equally to this work.}
\author{Florian~J.~Bennebroek~Evertsz'}
\thanks{These authors contributed equally to this work.}
\affiliation{QuTech and Kavli Institute of NanoScience, Delft University of Technology, 2600 GA Delft, The Netherlands}
\author{Gorm~O.~Steffensen}
\thanks{These authors contributed equally to this work.}
\affiliation{Departamento de F\'{i}sica Te\'{o}rica de la Materia Condensada, Universidad Aut\'{o}noma de Madrid, Madrid, Spain}
\affiliation{Condensed Matter Physics Center (IFIMAC), Universidad Aut\'{o}noma de Madrid, Madrid, Spain}
\affiliation{Instituto de Ciencia de Materiales de Madrid (ICMM), \\ Consejo Superior de Investigaciones Científicas (CSIC), \\ Sor Juana Inés de la Cruz 3, 28049 Madrid, Spain}
\author{Tom~Dvir}
\email{Currently at: Quantum Machines, Tel Aviv, Israel}
\author{Grzegorz~P.~Mazur}
\author{David~van~Driel}
\author{Nick~van~Loo}
\author{Jan~Cornelis~Wolff}
\affiliation{QuTech and Kavli Institute of NanoScience, Delft University of Technology, 2600 GA Delft, The Netherlands}
\author{Erik~P.A.M.~Bakkers}
\affiliation{Department of Applied Physics, Eindhoven University of Technology, 5600 MB Eindhoven, The Netherlands}
\author{Alfredo~Levy~Yeyati}
\affiliation{Departamento de F\'{i}sica Te\'{o}rica de la Materia Condensada, Universidad Aut\'{o}noma de Madrid, Madrid, Spain}
\affiliation{Condensed Matter Physics Center (IFIMAC), Universidad Aut\'{o}noma de Madrid, Madrid, Spain}
\affiliation{Instituto Nicol\'{a}s Cabrera, Universidad Aut\'{o}noma de Madrid, Madrid, Spain}
\author{Leo~P.~Kouwenhoven}
\email{l.p.kouwenhoven@tudelft.nl}
\affiliation{QuTech and Kavli Institute of NanoScience, Delft University of Technology, 2600 GA Delft, The Netherlands}

\date{\today}

\begin{abstract}
Detection and control of Andreev Bound States (ABSs) localized at semiconductor-superconductor interfaces are essential for their use in quantum applications. Here we investigate the impact of ABSs on the supercurrent through a Josephson junction containing a quantum dot (QD). Additional normal-metal tunneling probes on both sides of the junction unveil the ABSs residing at the semi-superconductor interfaces. Such knowledge provides an ingredient missing in previous studies, improving the connection between theory and experimental data. By varying the ABS energies using electrostatic gates, we show control of the switching current, with the ability to alter it by more than an order of magnitude. Finally, the large degree of ABS tunability allows us to realize a three-site ABS-QD-ABS molecule (Andreev trimer) in which the central QD is screened by both ABSs. This system is studied simultaneously using both supercurrent and spectroscopy.
\end{abstract}

\maketitle

\section*{Introduction}

Quantum dots (QDs) confine electrons into orbitals with discrete energies, similar to individual atoms~\cite{kouwenhoven2001few}. They find countless applications as sensors~\cite{Lu2003May,zhu2015photoluminescence}, light sources~\cite{michler2000quantum, deArquer2021Aug} or qubits~\cite{Hanson.2007, Burkard2023Jun}. Superconductors, on the other hand, feature an attractive pairing between electrons, condensing them into a sea of Cooper pairs~\cite{tinkham2004introduction}. One consequence of this pairing is the ability to carry supercurrent: zero-resistance transport of electron pairs. Supercurrent can flow even if two superconducting leads are connected by a thin insulating material or a weak link. Such a system is known as a Josephson junction and forms the core component behind superconducting qubits~\cite{clarke2008superconducting, Kjaergaard2020Mar} and superconducting diodes~\cite{ando2020observation}.

Using semiconducting QDs as weak links in Josephson junctions combines the precise orbital tunability of QDs with the quantum coherent properties of superconductors, resulting in substantial control over the supercurrent~\cite{jarillo2006quantum, vanDam2006supercurrent, jorgensen2007critical, katsaros2010hybrid,szombati2016josephson} and facilitating cQED operation~\cite{bargerbos2022singlet, Bargerbos2023Aug}. Intriguingly, QDs can also hybridize with a superconductor. For example, when hosting an odd number of electrons a QD acts as a localized spin $1/2$, which becomes screened by quasiparticles at stronger coupling, resulting in a spin-less Yu-Shiba-Rusinov groundstate~\cite{maurand2012first, lee2012zero, pillet2013tunneling, Jellinggaard.2016}. Control of this interaction allows for tuning of both the groundstate composition and the spectrum; an interesting lever for Andreev Spin Qubits~\cite{Padurariu2010Apr,hays2021coherent, Pita-Vidal2023Aug, Pita-Vidal2023Jul}, Kitaev chains~\cite{dvir2023realization, tsintzis2022creating, tenHaaf2023engineering} and the creation of larger superconducting molecules~\cite{Deacon2015Jul, Probst2016Oct, bouman2020triplet, saldana2020two}. 
However, typical devices are prone to the formation of accidental QDs or localized ABSs due to defects or impurities~\cite{pan2020physical, prada2020andreev, Valentini2021Jul}. Such states are hard to characterize and control and are generally detrimental to device operation. Conversely, recent works have shown that gate-controlled ABSs are useful for tuning the coupling between sites in minimal Kitaev chains \cite{liu2022tunable, bordin2023tunable, zatelli2023robust, Liu2023Oct}, spurring further interest into measuring and manipulating them.

In this work, we investigate – theoretically and experimentally – the impact of ABSs on a QD-based Josephson junction, and distinguish its features from a S-QD-S junction. We show how gate-tuning of ABSs affects both the size and gate symmetries of the critical current; an important quantity for e.g. Andreev Spin Qubits, whose readout signal is proportional to it~\cite{Pita-Vidal2023Aug, Pita-Vidal2023Jul}. Furthermore, we study the influence of the ABS-QD tunnel coupling on the screening of the odd-parity charge sector, comparing measurements of zero-bias conductance and critical current. In this manner, we realize an ABS-QD-ABS Andreev trimer where the screening of the central QD is facilitated by ABSs in both leads, highlighting their potential use in realizing larger chains and artificial molecules. 

\section*{Device}

\begin{figure}[ht!]
    \centering
    \includegraphics[width=0.5\columnwidth]{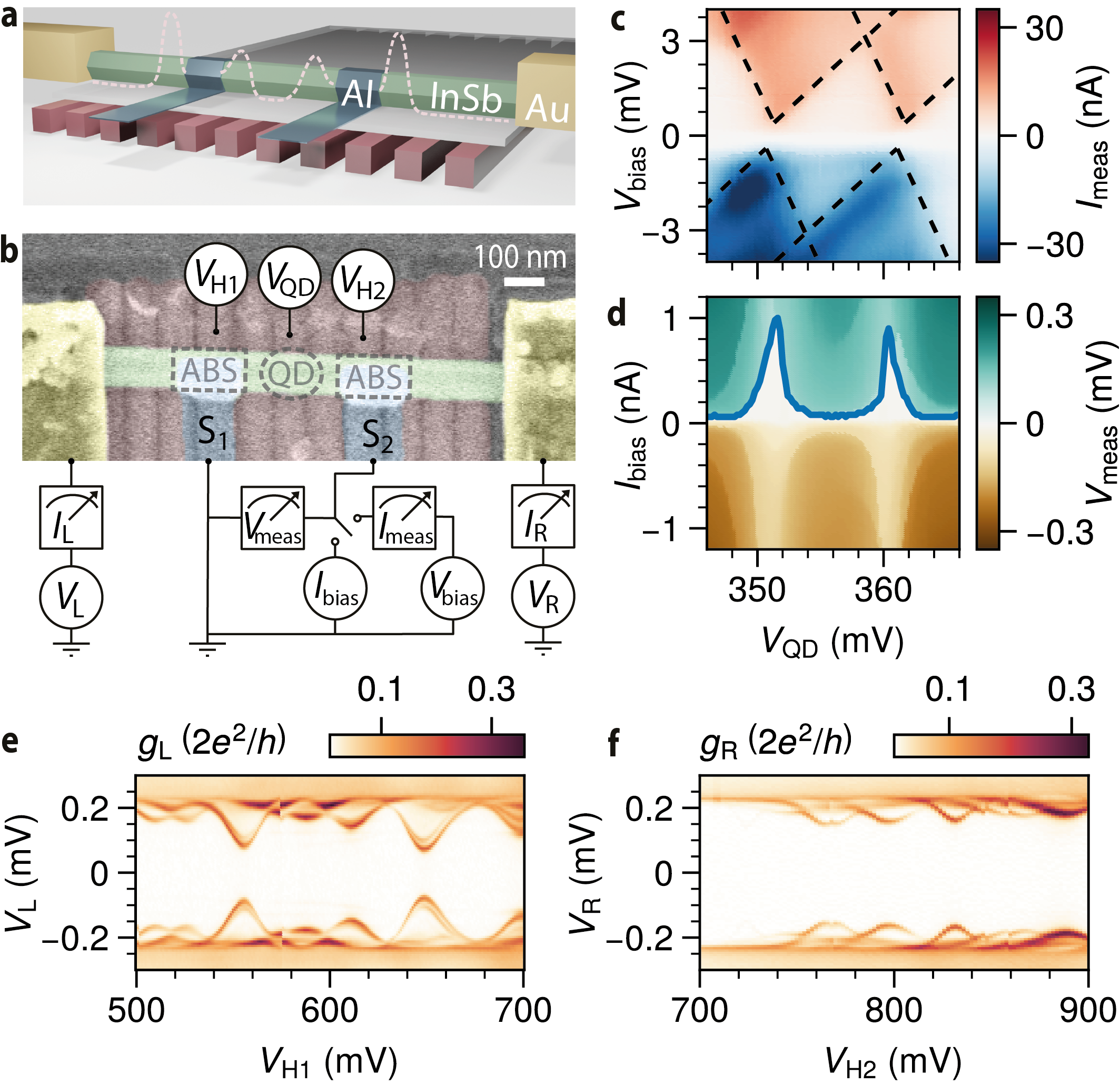}
    \caption{\textbf{a.} Schematic illustration of the device.
    \textbf{b.} A false-coloured SEM of the device and schematic of the measurement setup. 
    \textbf{c.} $\Imeas$ as a function of $\VQD$ and $\Vbias$ showing a single Coulomb diamond. The dashed lines show the fit to a QD model with charging energy $U = \SI{2.7}{meV}$ and lever arm $\alpha = 0.26$. \textbf{d.} $\Vmeas$ as a function of $\VQD$ and $\Ibias$. The blue line identifies the switching current, $\Isw$, where the Josephson junction transitions from a superconducting to a resistive regime. \textbf{e, f.} Tunneling spectroscopy of the left and right hybrids respectively, where $\gL \equiv \frac{d\IL}{d\VL}$ and $\gR \equiv \frac{d\IR}{d\VR}$ are measured with standard lockin techniques.}
    \label{fig:1}
\end{figure}

A tunable Josephson junction can be realized by contacting a semiconducting nanowire with two superconducting leads~\cite{vanDam2006supercurrent, lee2012zero, pillet2013tunneling, saldana2018supercurrent}. Here we start from an InSb nanowire with two Al contacts deposited using the shadow-wall lithography technique~\cite{Heedt.2021}. In addition, we introduce on each side of the device two normal leads, that will serve as probes. Below the nanowire, separated by a thin dielectric, an array of bottom gates can define an electrostatic potential along the device, as Fig.~\ref{fig:1}a illustrates. Two inner tunneling barriers form a QD and two additional outer barriers are tuned to separate each superconducting contact from the external normal contacts. These outer barriers turn both gold contacts into tunneling probes, which are connected to corresponding voltage sources and current meters ($\VL$, $\IL$ and $\VR$, $\IR$).

The Josephson junction in the middle of our device is connected to a flexible circuit that enables setting either a voltage bias or a current bias across the junction. Fig.~\ref{fig:1}b shows that while the left Al contact is always grounded, the right one is connected to a switch between either a voltage source with a current meter ($\Vbias$, $\Imeas$) or a current source with a voltage meter ($\Ibias$, $\Vmeas$). See the Methods for further nanofabrication and circuit details.
Panel c of Fig.~\ref{fig:1} reports a voltage bias measurement characterizing the QD's Coulomb blockade diamonds. In panel d, instead, we apply a current bias and observe a $\Vmeas=0$ region, which is identified as the DC supercurrent regime. The transition from zero to finite voltage is marked by a blue line, with the associated current bias values denoted as $\Isw$, the switching current. As previously demonstrated in literature, $\Isw(\VQD)$ depends sensitively on the QD gate voltage. The device behaves like a supercurrent transistor: $\Isw$ is maximal at the QD charge degeneracy points, while it is suppressed when the QD energy levels are off-resonance~\cite{jarillo2006quantum, vanDam2006supercurrent}.

The novelty of our device emerges from panels e and f of Fig.~\ref{fig:1}, which show tunneling spectroscopy measurements performed from the left and the right normal-metal leads yielding differential conductance $\gL$ and $\gR$, respectively. They reveal the density of states of the hybrid segments underneath each Al contact~\cite{tinkham2004introduction}. Both hybrid segments feature ABSs that disperse as a function of the voltage of the gates underneath ($\VHi$, $\VHii$). To understand their implications, we introduce in the following section a minimal three-site model, considering a single ABS in the left hybrid, a single QD orbital in the centre and a single ABS in the right hybrid.

\section*{Model} 
\begin{figure}[ht!]
    \centering
    \includegraphics[width=0.5\columnwidth]{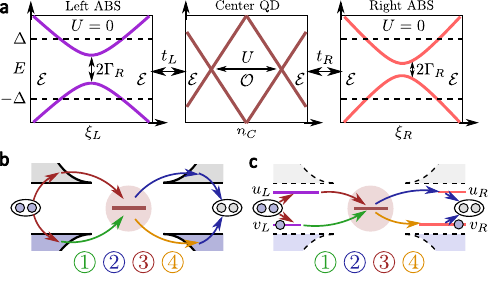}
    \caption{\textbf{a.} Schematic of the ABS-QD-ABS model. Smaller sketches resembling this schematic are used in the following figures to indicate gate settings. Here the scale $\Delta$ illustrates the position of the BCS continuum neglected in the model. The central plot is shown in scale of $U$. Symbols $\mathcal{E}$ and $\mathcal{O}$ indicate even and odd groundstate parity respectively. \textbf{b-c.} Sketches of $4^\mathrm{th}$ order contributions to the critical current, $I_c$, for a S-QD-S model (\textbf{b}) and an ABS-QD-ABS model (\textbf{c}). The numbers below indicate the ordering of the dominant $4^\mathrm{th}$ order process.
    }
    \label{fig:2}
\end{figure}

The left and right ABSs are modelled as single levels with negligible interaction ($U = 0$) coupled to BCS superconductors by couplings $\Gamma_{L/R}$ in the atomic limit ($\Delta \rightarrow \infty$)~\cite{Bauer.2007}. Both are tunnel coupled to a central QD with a large charging energy, $U = 10\Delta$, yielding the following Hamiltonian:
\begin{align}
H &= H_{ABS} + H_{D} + H_T \label{eq:FullH} \\
H_{ABS} &= \sum_{j=L/R}\left[\xi_j n_j + \Gamma_j d^\dagger_{j\uparrow }d^\dagger_{j\downarrow}+\text{h.c.}\right] \\
H_D &= \frac{U}{2}\left(n - n_C\right)^2 \\
H_T &= \sum_{j=L/R}\sum_\sigma t_{j} d^\dagger_{C\sigma}d_{j\sigma} + \text{h.c.}
\end{align}
Here, $\xi_{L/R}$ are the single-level energies, with $n_{L/R}$ and $d^\dagger_{L/R\sigma}$ denoting the corresponding number and creation operator, giving the ABS energies $E_{L/R} = \sqrt{\xi_{L/R}^2 + \Gamma_{L/R}^2}$. The number and creation operator of the central dot are denoted $n$ and $d^\dagger_{C\sigma}$ respectively, with $n_C$ describing the electrochemical potential controlled by gate. Lastly, the QD is tunnel coupled to the ABSs by couplings $t_{L/R} = |t_{L/R}|\exp\left[i\phi_{L/R}\right]$, with the phase drop across the junction characterized by the difference $\phi = \phi_L - \phi_R$. 
A sketch of the model is depicted in Fig.~\ref{fig:2}a. This model neglects both the detailed structure of the ABSs, e.g. multiple orbitals, when tuning away from resonance, and any screening of the QD due to a direct coupling to the BCS density of states of the leads~\cite{pillet2013tunneling, Lee.2014}, capturing only the screening stemming from the coupling to ABSs. Consequently, the validity of the model is limited to where ABSs are tuned close to resonance. More details can be found in the Methods.

For low QD-ABS coupling, the critical current $I_c$, of this three-site model can be understood in terms of the lowest ($4^\mathrm{th}$) order perturbation theory. Qualitatively, this yields similar results as a S-QD-S junction with alternating $0$ and $\pi$ phases corresponding to even and odd QD occupation~\cite{vanDam2006supercurrent, Martin-Rodero2011Dec}. Peaks of $I_c$ occur close to QD parity transitions, with the odd parity $I_c$ reduced compared to the even.
However, $\xi_{L/R}$ can be tuned in the ABS-QD-ABS system and has the effect of both changing the ABSs energy and the particle-hole coherence factors $u_{L/R}$ and $v_{L/R}$, where $u_{L/R}^2 = \frac{1}{2} \left( 1 + \frac{\xi_{L/R}}{E_{L/R}}\right)$ and $u_{L/R}^2 + v_{L/R}^2 = 1$~\cite{Bauer.2007} (see also Fig.~\ref{fig:3}b). This strongly enhances $I_c$ close to the parity transition, $n_C = 0.5$ or $n_C = 1.5$, with leading term,
\begin{equation}
I_{c,peak} \approx \frac{2e}{\hbar}\frac{\Gamma_L\Gamma_R |t_L|^2 |t_R|^2}{E_L^2 E_R^2 (E_L+E_R)},
\end{equation}
when approaching the parity transition from the even parity side. Here, one power of the $(E_L E_R)^{-1}$ dependence stems from the virtual excitation of the ABSs, and the other from the particle-hole factors since $I_c\propto |u_L v_L| |u_R v_R| = \Gamma_L \Gamma_R / (4E_L E_R)$~\cite{florian2023}. The leading $4^\mathrm{th}$ order processes are illustrated in Fig.~\ref{fig:2}b-c for an S-QD-S and an ABS-QD-ABS system respectively. Lastly, for stronger couplings, we use numerical diagonalization of eq.~(\ref{eq:FullH}), the results of which are shown in theory plots throughout the paper. This also captures ABS-QD hybridization, which depends on $u_{L/R}$, $v_{L/R}$ and $n_C$.     

\section*{Supercurrent control}

\begin{figure}[ht!]
    \centering
    \includegraphics[width=\columnwidth]{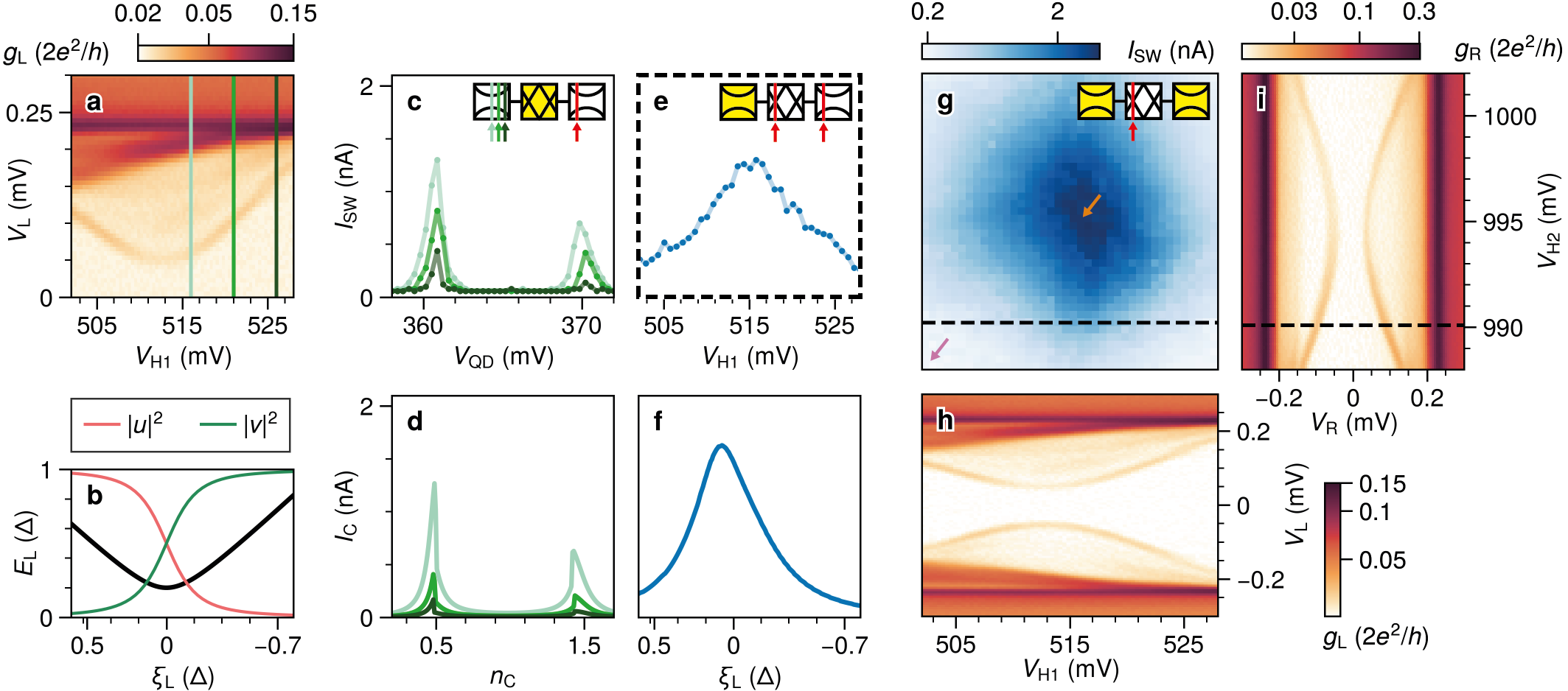}
    \caption{\textbf{a.} $\gL$ as a function of $\VHi$, showing the energy dispersion of a single ABS localised in the left hybrid segment. \textbf{b.} ABS energy and square of the particle-hole components $u_L$ and $v_L$ as predicted by the theory model. \textbf{c.} $\Isw$ as a function of $\VQD$ ranging over a single QD orbital and colour-coded to the vertical line cuts in panel a. \textbf{d.} $I_c$ predicted by the theory model. \textbf{e.} Maximum $\Isw$ of each line cut as a function of $\VHi$. \textbf{f.} Maximum $I_c$, as a function of $\xi_{\mathrm{L}}$. \textbf{g.} $\Isw$ as a function of $\VHi$ and $\VHii$. The black dashed line indicates the position along $\VHii$ at which the $\Isw$ in panel \textbf{e} is extracted. See Fig.~\ref{supp:4D} for extra details. \textbf{h.} $\gL$ as a function of $\VHi$, showing the same ABS as panel \textbf{a} but including its counterpart below the Fermi energy. \textbf{i.} $\gR$ as a function of $\VHii$, showing a single ABS localised in the right hybrid segment.  For both panels h and i, the QD gate is set next to the left QD resonance at $\VQD = \SI{358}{mV}$.}
    \label{fig:3}
\end{figure}

The theoretical model indicates that ABSs have an important effect on the supercurrent through the QD, which can be modulated by tuning the ABS excitation energy. To test their contribution, we focus in Fig.~\ref{fig:3}a on a single ABS, weakly coupled to the QD, and study its effects. 
In panel c, we measure $\Isw$ as a function of $\VQD$ at three different positions along $\VHi$, colour-coded to the vertical line cuts shown in panel a. We observe an overall increase in $\Isw$ as we move closer to the minimum energy of the ABS, which is reproduced by the model in panel d. In comparing theory to experiment we distinguish between measured switching current, $\Isw$, and theoretical critical current, $\Ic$, since $\Isw$ might be affected by circuit noise and thermal fluctuations~\cite{tinkham2004introduction}.

Alternatively, we may track the $\Isw$ peak value along one of the QD resonances (as detailed in Fig.~\ref{supp:4D}c). In Fig.~\ref{fig:3}e we plot $\Isw$ (blue), tracing the left QD resonance, as a function of $\VHi$. We observe a maximum $\Isw$ of approximately $\SI{1.2}{nA}$ around the ABS minimum and a decline in $\Isw$ as we move away from this energy minimum. Notably, $\Isw$ is decreased as low as $\SI{0.2}{nA}$ when $E_{\mathrm{L}}$ approaches $\Delta$, suggesting that most of the supercurrent is mediated via the ABS and not via continuum states at the gap edge of the superconductor. This is supported by the model, as presented in panel f, which does not include these continuum states and yields a similar decrease. For a detailed explanation of the fitting procedure see Fig.~\ref{supp:fitting-procedure}.

Thus far, we kept the ABS in the right hybrid segment, presented in panel i, fixed at $\VHii = \SI{990}{mV}$. In Fig.~\ref{fig:3}g, we present $\Isw$ as a function of both ABS gates, tracking the $\Isw$ peak of the left QD resonance (see Fig.~\ref{supp:4D} for details). The black dashed line illustrates the $\VHii$ set point along which we extract the data for panels c and e. In the corners of panel d, both ABSs are off-resonance, resulting in a minimal $\Isw$ of about $\SI{0.2}{nA}$ (pink arrow). Along the sides of the panel, a single ABS reaches a resonance resulting in an increase of $\Isw$ to around $\SI{1}{nA}$. In the middle of the panel, both ABSs are on resonance resulting in a maximum enhancement of $\Isw$ up to $\SI{2.58}{nA}$ (orange arrow).
Controlling the ABSs, we can modulate $\Isw$ by over an order of magnitude.

\section*{Andreev trimer}

\begin{figure}[ht!]
    \centering
    \includegraphics[width=\columnwidth]{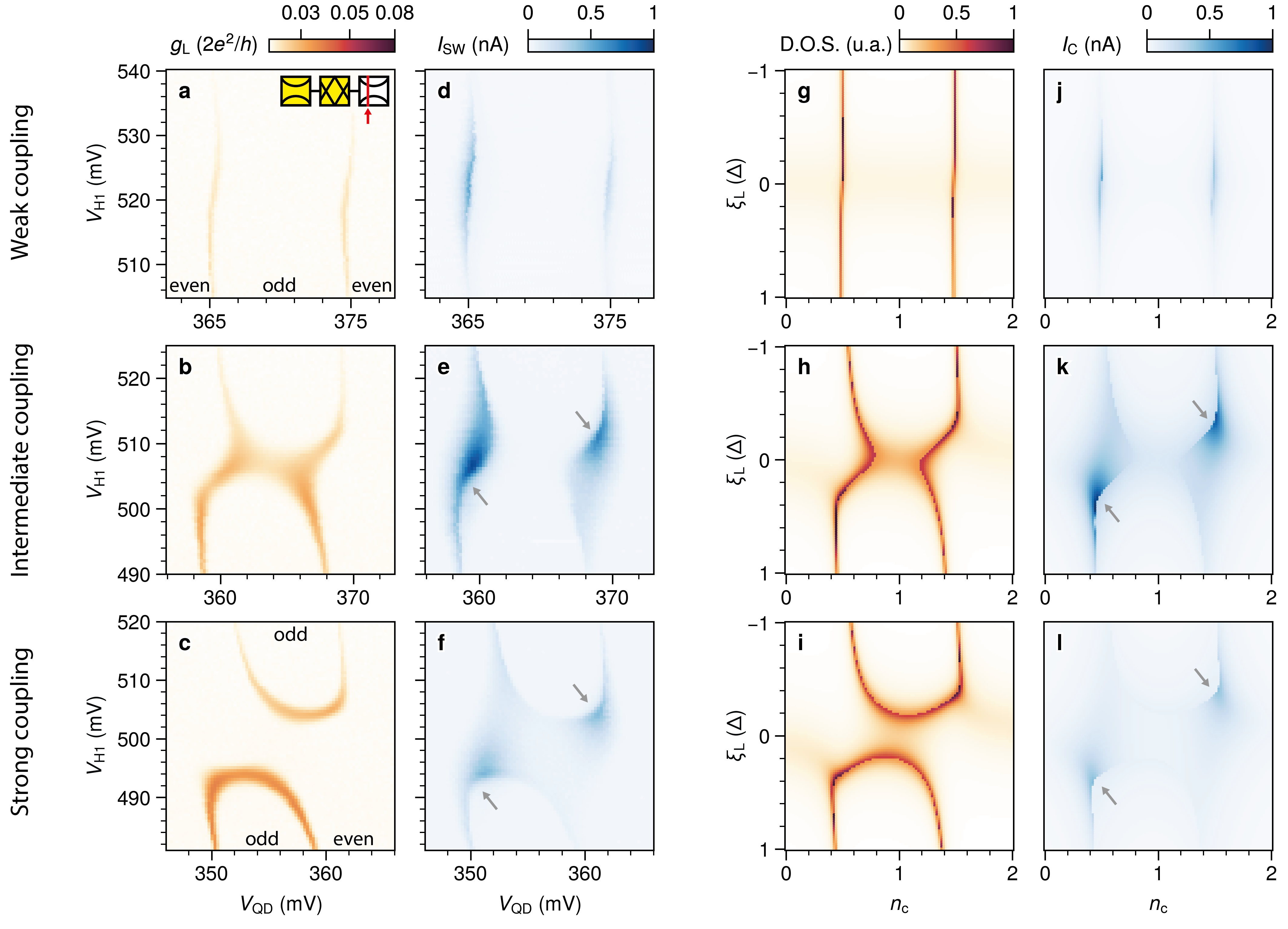}
    \caption{\textbf{a-c.} Zero-bias conductance $\gL$ as a function of $\VQD$ and $\VHi$ for weak, intermediate and strong ABS-QD hybridization. \textbf{d-f.} 2D maps of the switching current corresponding to the same gate range of panels \textbf{a-c}.  \textbf{g-i.} Theory simulation of the zero-energy density of states using the Lehmann representation (see Methods). \textbf{j-l.} Simulated critical current using our minimal three-site model. Panels (\textbf{g}, \textbf{j}), (\textbf{h}, \textbf{k}) and (\textbf{i}, \textbf{l}) share the same model parameters.}
    \label{fig:4}
\end{figure}

After demonstrating switching current control, we turn our attention to the physics of an ABS-QD-ABS molecule. Conceptually, this setup is reminiscent of a S-QD-S junction.
However, while the screening states of the S-QD-S junction are of a complicated Kondo-like nature~\cite{Martin-Rodero2011Dec, Meden2019Feb}, the ABS-QD-ABS equivalents are simpler. Here, the half-filled QD binds to an excited ABS, gaining an exchange energy of order $\sim |t_{L/R}^2|/U$. If the exchange energy exceeds $\Gamma_{L/R}$, then the ground state is rendered into a molecular singlet state. To investigate this bonding, we make use of the supercurrent as a characterization tool. First, we fix the electrochemical potential of the right ABS and study the coupling between the left ABS and the QD looking at the ABS-QD charge stability diagrams of Fig.~\ref{fig:4}. This can be measured via either tunneling spectroscopy from the normal-metal probes on the sides of our device or via supercurrent through our Josephson junction. In the first case, the charge degeneracy points are identified by zero-bias conductance peaks and in the second one by switching current peaks. 

In our device, we can tune the ABS-QD hybridisation from weak to strong, while the ABS spectra are continuously monitored. %This leaves $t_L$ and $t_R$ as the only two free modelling parameters to fit between different tunings. 
When the coupling is weak (panels a,d), the QD charge degeneracy points are barely shifted along $\VQD$ as we sweep over a single ABS gate. As a result, the odd parity sector separating the even parity ones remains roughly equal in width. In the intermediate coupling case (panels b,e), the QD resonances are visibly modulated and the odd parity sector is reduced. In the strong case (panels c,f), the topology of the ABS-QD charge stability diagram is changed, due to the screening of QD spins by the ABS~\cite{grove2018yu, saldana2020two}. As a result, the system no longer transitions to an odd parity ground state when the ABS is on resonance. All coupling regimes are accurately reproduced by our model in both conductance and supercurrent simulations (panels g-l). Other parameters than $t_L$ and $t_R$ are estimated from independent measurements such as ABS spectroscopy and QD Coulomb diamonds (see Fig.~\ref{supp:fitting-procedure}). In particular, since here we set the right ABS more and more off-resonance as the coupling gets stronger to get comparable $\Isw$ colorbars, the extracted $\xi_R$ values vary from row to row. The summary of all extracted and fitted model parameters is reported in Table~\ref{tab:summary}.

In the weak coupling regime, the measured switching current is mostly captured by the 4$^\mathrm{th}$ order perturbation theory; the presence of ABSs leads to a quantitative difference in $\Ic$, due to lower energy excitations, $E_{L/R}$, mimicking a smaller $\Delta$.
In the intermediate and strong coupling regimes, the presence of ABSs leads to a qualitative difference as well. For instance, the grey arrows in Fig.~\ref{fig:4}e,f,k,l highlight strong asymmetries in the switching current peak heights. Such asymmetries are due to the $u$ and $v$ components of the ABSs. Approaching e.g. $n_C \sim 0.5$, the QD-ABS hybridization is strongest for an ABS with $|u|\gg |v|$ as then both the ABS and the QD are most easily excited by the addition of an electron. This stronger hybridization results in a higher $\Ic$, as shown also in Fig.~\ref{fig:SupTheory1}.
Switching current peak asymmetries were previously attributed to multiple QD orbitals~\cite{vanDam2006supercurrent}. Here we propose an additional possibility for hybrid devices: asymmetries explained by the coherence factors of subgap ABSs.

\begin{figure}[ht!]
    \centering
    \includegraphics[width=0.56\columnwidth]{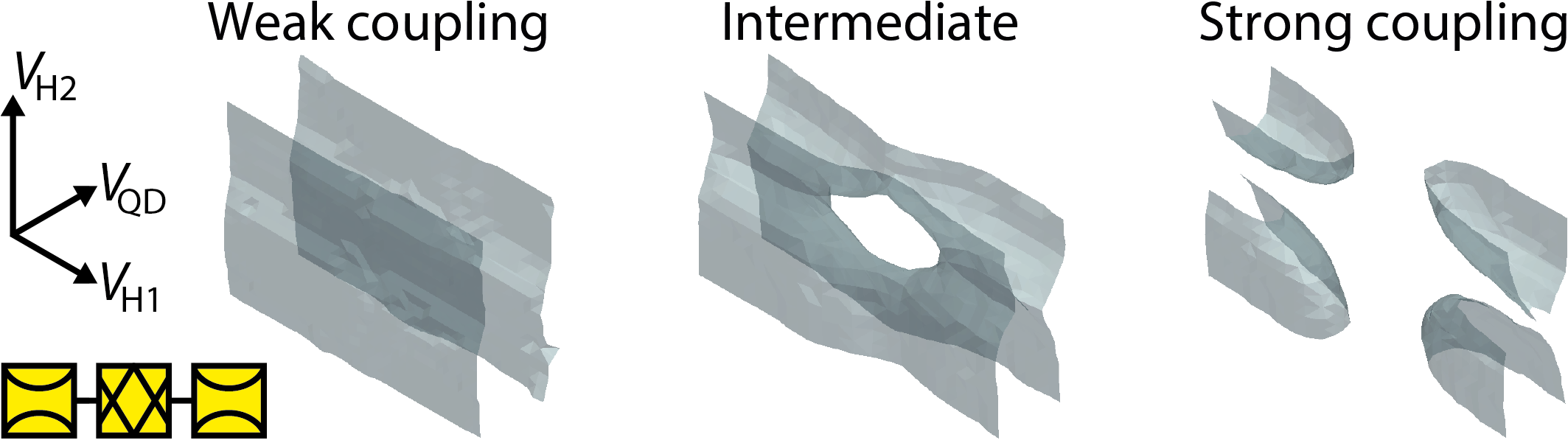}
    \caption{Measured 3D charge stability diagrams for weak, intermediate and strong coupling between the QD and the neighbouring ABSs. See Fig.~\ref{supp:3D-CSD} for the measurement details.}
    \label{fig:5}
\end{figure}

Finally, we turn our attention to the full three-site Andreev molecule by varying the right ABS as well. Fig.~\ref{fig:5} shows 3D charge stability diagrams extracted from zero-bias conductance measurements (see also Fig.~\ref{supp:3D-CSD}). In the weak coupling regime, the charge degeneracy manifolds are two parallel planes, isolating an odd-parity region between two even-parity ones; here, varying $\VQD$ can always switch the ground state parity, regardless of the ABS tuning, indicating the independence of three components. In the intermediate coupling regime, the situation is different, as can be appreciated by the different topology of the charge degeneracy manifold, which presents a hole connecting the two even-parity regions. In this regime, it is only when both ABSs are simultaneously at their energy minimum that the doublet QD state can be screened to a singlet, while a single ABS at minimum energy cannot fully screen the odd parity sector. This shows that the two ABSs can cooperate in the screening of the QD spins, expanding the regions where the system has an even-parity ground state.
The even-parity regions expand even further in the strong coupling regime, where the charge degeneracy manifold topology is changed once more. Here a single ABS set on resonance is able to fully screen the odd parity QD state, as observed by the odd-parity domes only being present in the four corners of the diagram where both ABS are tuned away from resonance.

\begin{figure}[ht!]
    \centering
    \includegraphics[width=0.5\columnwidth]{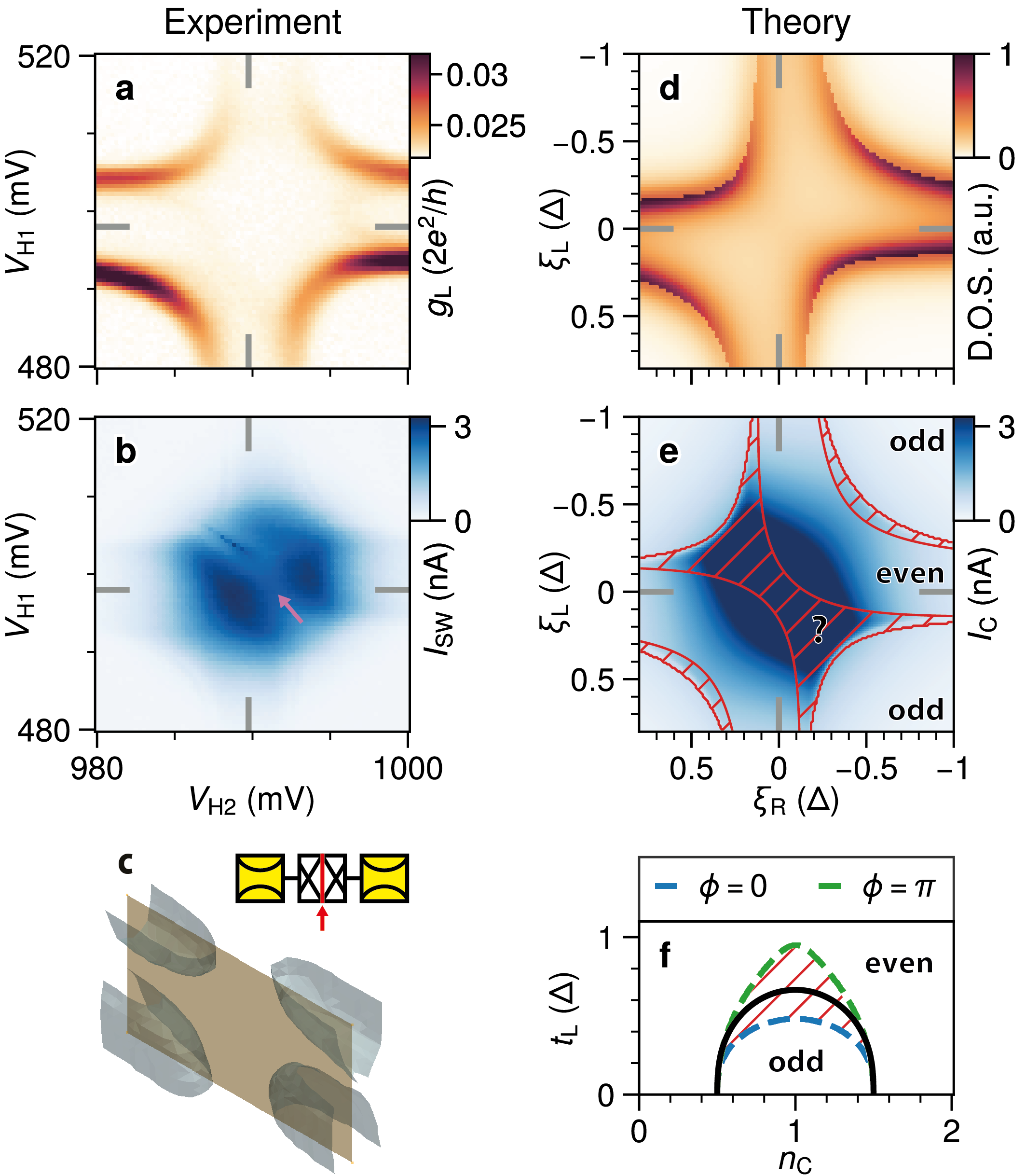}
    \caption{\textbf{a.} Left zero-bias conductance $\gL$ as a function of $\VHi$ and $\VHii$ in the strong coupling regime. Grey lines indicate the positions of the charge neutrality points of both ABSs. \textbf{b.} A 2D switching current map. In both panel \textbf{a} and \textbf{b}, $\VQD$ is placed between the two QD resonances as panel \textbf{c} indicates. \textbf{d.}  Theory simulation of the zero-energy density of states using the Lehmann representation. \textbf{e.} Simulated critical current using our minimal three-site model. Panels \textbf{d} and \textbf{e} share the same model parameters. \textbf{f.} A phase space diagram indicating the ground state transitions of the system at $\phi = 0$ (blue) and $\phi = \pi$ (green) for parameters of d and e with $t_L = 0.7$ and $t_R=0.8$. For comparison, in black the ground state transitions are illustrated assuming a single ABS. The red-shaded areas of panels \textbf{e} and \textbf{f} indicate the region where the groundstate parity may depend on phase.}
    \label{fig:6}
\end{figure}

To appreciate the effect on the supercurrent of such parity transitions, we focus in Fig.~\ref{fig:6} on the strong coupling regime. Keeping the QD gate fixed between the two resonances, we study the system as a function of both ABS gates, as indicated in panel c. Panel a shows the measured charge stability diagram, while panel b shows the corresponding $\Isw$ map.
The theory counterpart is presented in panels d and e. Here, both maps reproduce the experimental features, apart from a discrepancy between $\Isw$ in b and $\Ic$ in e at the centre when both ABSs are tuned to resonance. We speculate that this discrepancy stems from the possibility of groundstate transitions as a function of phase difference, $\phi$. This originates in the phase dependence of the screening. For $\phi \approx 0$, the two ABSs cooperate in screening, thereby decreasing the area of the odd doublet state compared to the effect of a single ABS, as seen in panel f. For $\phi = \pi$, however, the two ABSs instead compete, expanding the range of the odd doublet~\cite{Rozhkov1999Mar, Pavesic2023Apr}. The areas where these transitions with $\phi$ occur are shown with red overlays in e and f, and qualitatively match with the area of discrepancy between b and e. More details on current phase relations are shown in Fig.~\ref{fig:SupTheory2}, and its influence on the D.O.S in Fig.~\ref{supp:phase}. The mechanism that causes premature switching when groundstate transitions can occur with $\phi$ is beyond the scope of this work and will be investigated in a future study. We note that this discrepancy solely exists in the strong coupling regime, as can be seen in Fig.~\ref{supp:YSR-matrix}. Future works incorporating SQUIDs could unveil the interesting $\phi$-dependence of this regime. 

\section*{Conclusion}

In summary, we realised a QD embedded into a Josephson junction with additional side probes revealing neighbouring ABSs. These ABSs are shown to be the primary carrier of supercurrent, with measured switching currents matching the predictions of a simple three-site model. This illustrates the crucial role of controlling and detecting localized ABSs in semiconductor-superconductor hybrid devices.
Furthermore, by tuning couplings and ABSs we have demonstrated that the system effectively behaves as an Andreev trimer, whose charging diagram can be fully characterized via either supercurrent or normal probe measurements. This additionally exemplifies how ABS tuning can be done via supercurrent in long nanowire-based Kitaev chains, for which the normal probes would be further away from central ABSs \cite{tsintzis2023roadmap, miles2023kitaev, bordin2024crossed}. Besides that, our study sets the ground for future works on Josephson junction devices with increased complexity, including longer Andreev molecules predicted to modulate the supercurrent non-locally~\cite{kocsis2023strong} and complex Andreev spin qubit devices~\cite{Pita-Vidal2023Jul}.  

%TC:ignore


\section*{Acknowledgements}
This work has been supported by the Dutch Organization for Scientific Research (NWO) and Microsoft Corporation Station Q. 
We wish to acknowledge Isidora Araya Day, Anton Akhmerov and Jens Paaske for useful discussions and Ghada Badawy and Sasa Gazibegovic for the nanowire growth. GS and ALY acknowledge EU through FET-Open project ANDQC and Spanish AEI through grant TED2021-130292B-C43. 

\section*{Author contributions}

AB, JCW, and DvD fabricated the device. GOS and ALY developed the theoretical model and performed numerical simulations. FJBE and AB performed the electrical measurements with help from GPM and NvL. AB and TD designed the experiment. EPAMB provided the nanowires. LPK supervised the project. AB, FJBE, GOS, ALY and LPK prepared the manuscript with input from all authors.

\section*{Data availability}
All raw data in the publication and the analysis code used to generate figures are available at 
\url{https://doi.org/10.5281/zenodo.10711820}. 

\clearpage
\pagebreak
\newpage

\renewcommand\thefigure{ED\arabic{figure}}
\setcounter{figure}{0}

\section*{Methods}

\subsection*{Nanofabrication}

The nanofabrication process is identical to what is reported in ref.~\cite{bordin2024crossed}. An InSb nanowire is deposited with a micromanipulator on top of pre-patterned bottom gates (3/\SI{17}{nm} of Ti/Pd). We note that while our device has 11 bottom gates as in~\cite{bordin2024crossed}, only 7 of them are necessary for the experiment of the present manuscript: 3 central gates are used to form a QD, 2 gates – one per hybrid – are used to control the ABSs and 2 more gates are used to form tunneling barriers between the superconducting contacts and the outer normal-metal ones. The remaining 4 gates are not used, they are held at a fixed positive voltage for the full duration of the experiment. This ensures that the corresponding portions of the nanowire are not pinched-off and always conduct.

A bi-layer dielectric deposited with ALD separates the gates from each other and from the nanowire (10/\SI{10}{nm} of Al$_2$O$_3$/HfO$_2$). The superconducting Al contacts are deposited with the shadow-wall lithography technique after removal of the native oxide on the surface of the nanowire via hydrogen cleaning~\cite{Heedt.2021}. Finally, 10/\SI{120}{nm} of Cr/Au contacts are deposited on the two sides of the device with standard e-beam lithography after the removal of the native oxide with Ar milling.

We note that most choices of thicknesses and materials are not critical. The essential requirements are the creation of a QD in a Josephson junction and the formation of ABSs, which are ubiquitous across diverse platforms~\cite{de2010hybrid}. In our specific material combination, the ABSs are particularly visible thanks to the otherwise hard gap of our InSb-Al hybrids~\cite{Heedt.2021}, they are isolated from each other thanks to the confining geometry, they can extend far below the Al energy gap thanks to the tunability of our semiconductor~\cite{vanLoo2023electrostatic} and they can be analyzed thanks to the normal-metal probes on either side of our device. We have empirically demonstrated the impact of such ABSs on the supercurrent, emphasizing that it should not be disregarded in any Josephson junction device defined in hybrid materials, including superconductors in combination with InSb, InAs, Si, Ge nanowires and 2DEGs, carbon nanotubes and others.

\subsection*{Electrical circuit}

\begin{figure}[ht!]
    \centering
    \includegraphics[scale=0.5]{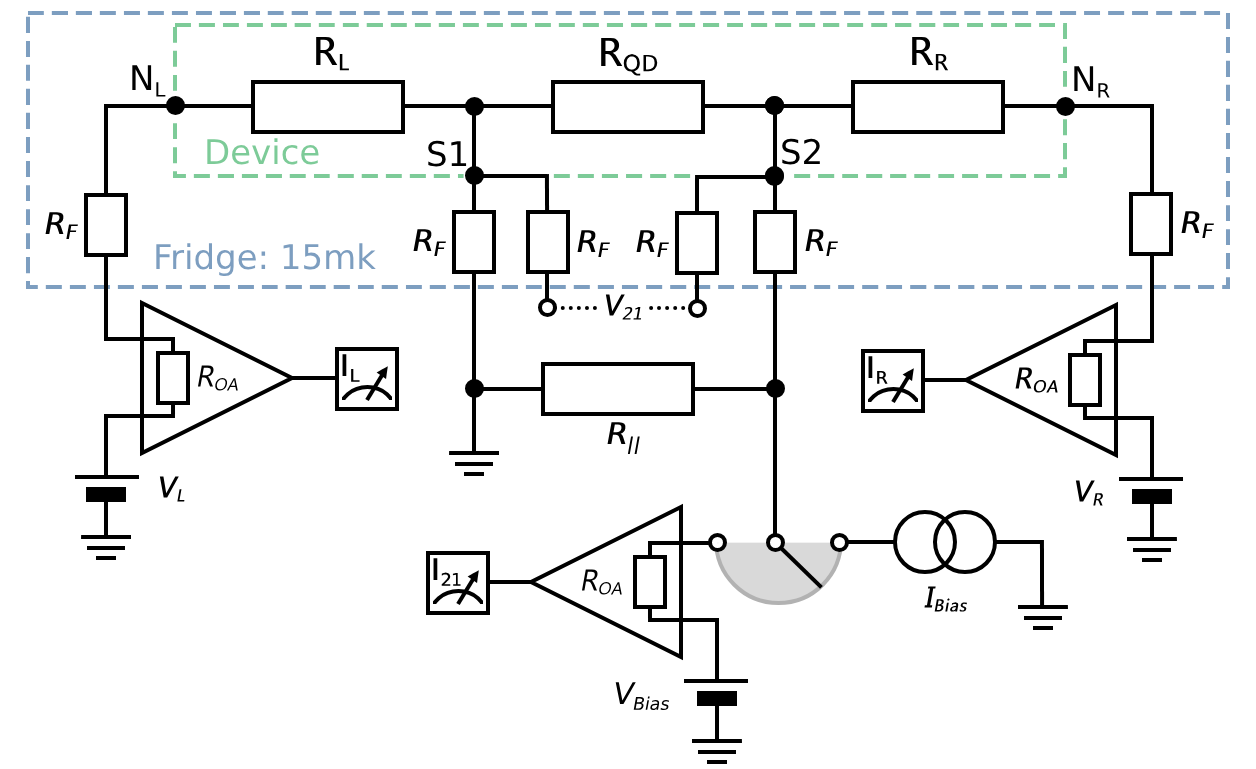}
    \caption{The measurement circuit. The device is enclosed by the green dashed line. $R_{\mathrm{L}}$, $R_{\mathrm{R}}$ and $R_{\mathrm{QD}}$ represent the tunneling barriers used for spectroscopy and the formation of the QD. The blue dashed line denotes the part of the circuit placed inside the dilution refrigerator at 15 mK. The device is connected to standard measurement equipment outside the fridge by fridge lines with a resistance $R_{\mathrm{F}}$. The switch allows for the transition between voltage and current driven experiments on the Josephson junction. The switch is presented open in this figure.}
    \label{fig:circuit}
\end{figure}

The device is placed inside a dilution refrigerator with a base temperature of $\approx \SI{15} {mK}$. It is connected to standard measurement equipment via fridge lines, resulting in an in-series resistance $R_{\mathrm{F}}$ at multiple points in the circuit as illustrated in Fig. \ref{fig:circuit}. 

The normal leads of the device, $\NL$ and $\NR$, are connected to independent voltage sources, $\VL$ and $\VR$, and current meters, $\IL$ and $\IR$. The use of operational amplifiers results in an additional series resistance $R_{\mathrm{OA}}$ added to $R_{\mathrm{F}}$. In all experiments presented in this work, $R_{\mathrm{L}}$, $R_{\mathrm{R}}$ and $R_{\mathrm{QD}} \gg R_{\mathrm{F}} + R_{\mathrm{OA}}$ such that their effect on transport experiments is negligible. We have therefore not corrected our data for voltage drops across these circuit resistances. Conductance measurements are performed using lock-in techniques with an AC excitation between $5$ and $\SI{10}{\mu V}$. 

The superconducting leads, $\SL$ and $\SR$, allow for both a voltage and current bias on the Josephson junction. In either configuration, $\SL$ is grounded and $\SR$ is driven. Transitioning between setups occurs via a switch as indicated in Fig. \ref{fig:circuit}. To prevent a potential build up of large voltages when the switch is open, an additional transport channel is created between $\SL$ and $\SR$ via a parallel resistor $R_{\mathrm{||}} = 10M\Omega$.

In the voltage bias setup, $\SR$ is connected to a voltage source $\Vbias$ and current meter $\Imeas$. This setup allows for characterization measurements of the QD in the form of Coulomb diamonds. For these measurements, $I_{21}$ is always corrected for the leakage current flowing through $R_{\mathrm{||}}$ and $R_{\mathrm{R}}$. To minimize the leakage current, we aim for $R_{\mathrm{||}}$ and $R_{\mathrm{R}} \gg R_{\mathrm{QD}}$. 

In the current bias setup, $\SR$ is connected to a current source $\Ibias$. The voltage over the junction $V_{\mathrm{21}}$ is measured using a four probe configuration to circumvent the fridge line resistances as illustrated in Fig. \ref{fig:circuit}. It is essential that, whilst in the supercurrent regime, all current flows through the Josephson junction and none through other transport paths such as $R_{\mathrm{||}}$ or $R_{\mathrm{R}}$ to accurately determine $\Isw$. This is ensured by the fact that $R_{\mathrm{QD}} = 0$ in the supercurrent regime. When the Josephson junction is resistive, this no longer holds and some of the applied current will leak away. We argue that this is not relevant since we are generally only interested in the supercurrent regime of the junction.

\subsection*{Theory}
In this section we elaborate on the theory used in the main text.
In general, the coupling of an interacting QD with a superconducting gap leads to the formation of YSR states which, in full treatment, requires techniques able to capture strong interaction, e.g. the Numerical Renormalization Group (NRG)~\cite{Bauer.2007}. In this paper we instead opt for a minimal model, capturing the dynamics of an ABS coupled to a QD qualitatively. We assume that both the left and right ABS can be described as a non-interacting resonant level ($U=0$) coupled to a superconducting lead with gap $\Delta$, shown in eq.~(\ref{eq:FullH}). The full Hilbert space of this Hamiltonian is $64\times 64$ and can be numerically diagonalized to obtain measureables. The supercurrent is given by,
\begin{equation}
I(\phi) = \frac{2e}{\hbar}\frac{\partial F(\phi)}{\partial \phi} \overset{T=0}{=} \frac{2e}{\hbar}\frac{\partial E_g(\phi)}{\partial \phi}
\end{equation}
with the systems free energy, $F$, being fully characterized by the groundstate energy, $E_g(\phi)$, at zero temperature, $T=0$, which is the limit used in the main paper. In our device geometry, we don't have active control over the phase, which therefore adjusts itself to satisfy the current-phase relationship $\phi(\Ibias, \Ic)$. The critical current is given by $I_c = \text{max}_{\phi}\left[I(\phi)\right]$ and, for a sinusoidal current phase relation, is either at $\phi = \pi/2$ or $\phi=3\pi/2$ for even and odd groundstate parity respectively~\cite{vanDam2006supercurrent}.

In the weak coupling limit, $t_j \ll U, \Gamma_j$, the critical current can also be analytically obtained via 4th order perturbation of the current operator in $t_L$ and $t_R$ \cite{Novotny2005Dec, saldana2018supercurrent},
\begin{align}
I_c(n=0) = \frac{e}{\hbar}\frac{\Gamma_L \Gamma_R|t_L|^2 |t_R|^2 }{E_L E_R}&\bigg[\frac{4}{\left(E_1 - E_0 + E_L\right)\left(E_2 - E_0\right)\left(E_1 - E_0 +E_R\right)} \\
+&\frac{2}{\left(E_1 - E_0 + E_L\right)\left(E_L + E_R\right)\left(E_1 - E_0 +E_R\right)}\bigg], \nonumber \\
I_c(n=1) = \frac{e}{\hbar}\frac{\Gamma_L \Gamma_R|t_L|^2 |t_R|^2 }{E_L E_R}&\bigg[\sum_{k=0,2}\frac{1}{\left(E_k - E_1 + E_L\right)\left(E_L+E_R\right)\left(E_k - E_1 +E_R\right)} \\
+ &\sum_{m=L,R}\frac{2}{\left(E_0 - E_1 + E_m\right)\left(E_L+E_R\right)\left(E_2 - E_1 +E_m\right)}\bigg], \nonumber \\
I_c(n=2) = \frac{e}{\hbar}\frac{\Gamma_L \Gamma_R|t_L|^2 |t_R|^2 }{E_L E_R}&\bigg[\frac{4}{\left(E_1- E_2 + E_L\right)\left(E_0 - E_2\right)\left(E_1 - E_2 +E_R\right)}
\\
+&\frac{2}{\left(E_1 - E_2 + E_L\right)\left(E_L + E_R\right)\left(E_1 - E_2 +E_R\right)}\bigg], \nonumber
\end{align}
where $E_0 = Un_C^2/2$, $E_1 = U(1-n_C)^2/2$, and $E_2 = U(2-n_C)^2$ are the central dot eigenenergies. Analytical expressions of $4^\textrm
{th}$ or higher order perturbation terms can be obtained using \texttt{pymablock}~\cite{Pymablock}.

Lastly, we evaluate the electron and hole component of the Lehmann representation from the eigenstates of eq.~(\ref{eq:FullH}) using,
\begin{align}
G_{je}^R(\omega) &= \sum_i\sum_{\sigma}\frac{|\bra{i}d^\dagger_{j\sigma}\ket{g}|^2}{\omega -E_i + E_g + i\eta} \\
G_{jh}^R(\omega) &= \sum_i\sum_{\sigma}\frac{|\bra{i}d_{j\sigma}\ket{g}|^2}{\omega -E_i + E_g + i\eta}
\end{align}
with $\ket{g}$ indicating groundstate and the sum of $i$ is over all other excitation's. A broadening, $\eta$, has been added for visibility. Using these, the local tunneling density of states of site $j$, as probed by a weakly coupled metallic lead~\cite{Dominguez2016Jan}, is given by,
\begin{equation}
\text{DOS}_j(\omega) = \text{Im}\left[G_{je}(\omega) + G_{jh}(-\omega)\right],
\end{equation}
which can be compared to measurements of the tunneling density of states using either the left or right metallic lead as a probe. Throughout the paper we plot DOS is arbitrary units, as the magnitude depends on unknown quantities such as initial density of states, etc. In general we find that the theory matches experimental data to qualitative degree across most gate settings, apart from the missing continuum not included in the model. An exception occurs in the regimes where one or both ABSs are brought far away from resonance, such that $E_{L/R} \approx \xi_{/R} > \Delta$. The primary contribution to both the screening of the central dot and supercurrent across it then stems from the gap, which is not captured by the model.

\begin{figure}[ht!]
    \centering
    \includegraphics[width=0.9\columnwidth]{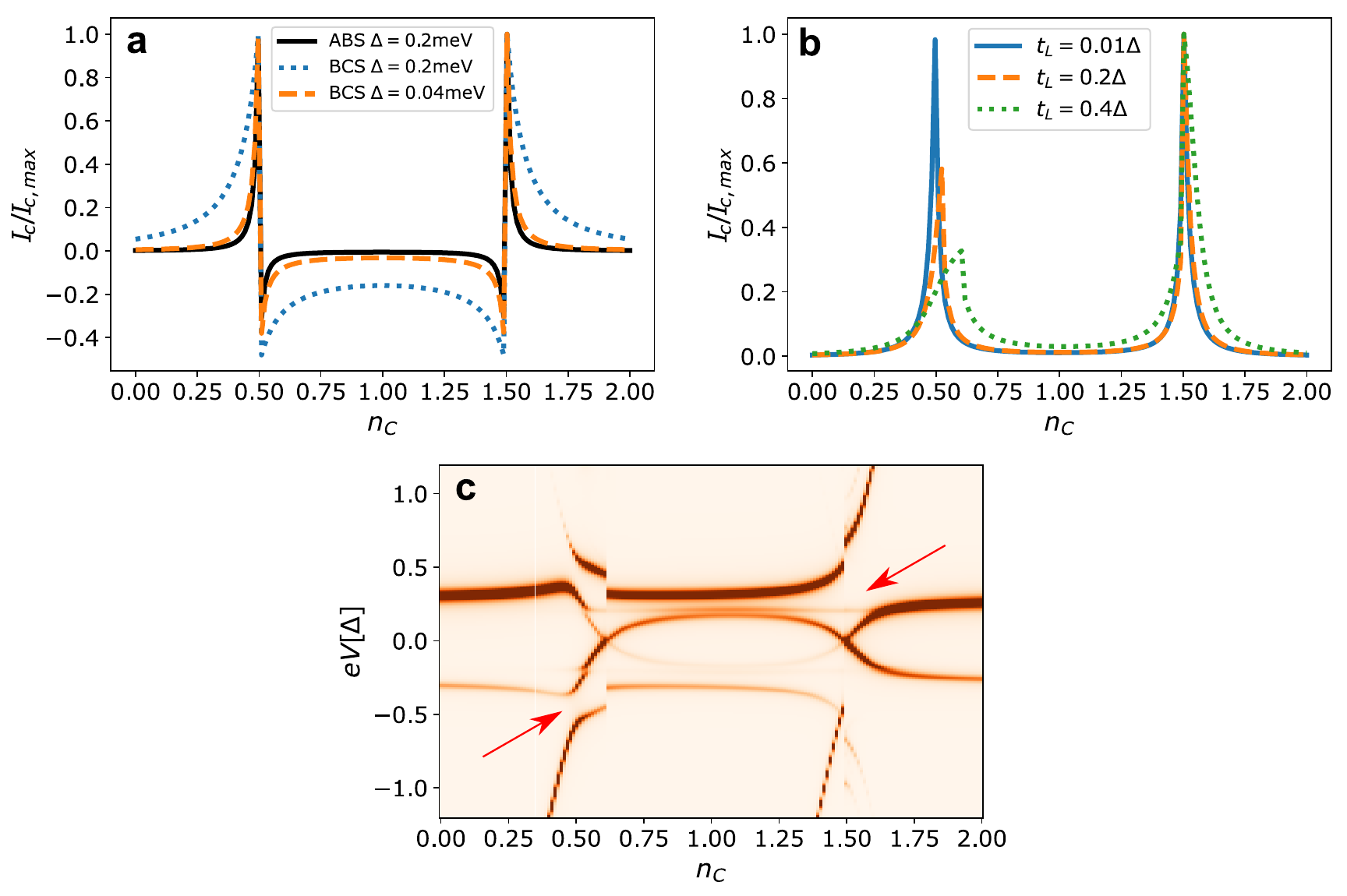}
    \caption{\textbf{a} Comparison between S-QD-S and ABS-QD-ABS in the low coupling limit, which for ABS is $t_L = t_R = 0.01\Delta$ and for BCS corresponds to coupling rate $\Gamma_{LS} = \Gamma_{RS} = 0.001\Delta$. Furthermore, $U = 2$~meV and $\xi_L = \xi_R = 0.0$ for all plots. Negative $I_c$ indicates $\pi$-phase. \textbf{b} Highlight of $I_c$ asymmetry at higher $t_L$ for finite detuning, $\xi_L = 0.2 \Delta$, with $\Delta = 0.2$~meV, $U=10\Delta$, $t_R = 0.1\Delta$, and $\xi_R = 0.0$. \textbf{c} DOS for identical parameters as \textbf{b} with $t_L = 0.4\Delta$.}
    \label{fig:SupTheory1}
\end{figure}

Next, we discuss various regimes of the model and its impact on experimental interpretation. In Fig.~\ref{fig:SupTheory1}a, we show $I_c$ at weak coupling such that the $4^\textrm{th}$ order approximation is valid, and compare it to the standard BCS $4^\textrm{th}$ order result \cite{saldana2018supercurrent}. In general for both models, if $U/\Delta$ is increased $I_c$ becomes more confined around the parity transitions. Choosing $\Delta = \Gamma_{L} = \Gamma_{R}$ yields very similar curves between ABS and BCS, supporting that in the low coupling regime the ABSs qualitatively acts as reduced gaps. In Fig.~\ref{fig:SupTheory1}b we show that for finite ABS detuning, $\xi_L = 0.2\Delta$, the critical current is initially symmetric between the two parity transitions for low coupling ($t_L = 0.01 \Delta$), but becomes asymmetric as coupling is increased. This highlights the breakdown of the 4th order expansion, which is always symmetric, and the appearance of QD-ABS hybridization. In Fig.~\ref{fig:SupTheory1}c we show the DOS for similar parameters, and highlight the different sizes of anti-crossings which correlate with the $I_c$ asymmetry. This relates to the coherence factors of the ABSs; for positive $\xi_L$ the hole component $\nu_L$, is amplified while at the $2$ to $1$ QD parity transition the QD is also most easily excited by the removal of an electron. At the $0$ to $1$ transition there is a mismatch as the QD is most easily excited by the addition of an electron, and so the hybridization is smaller.

\begin{figure}[ht!]
    \centering
    \includegraphics[width=0.9\columnwidth]{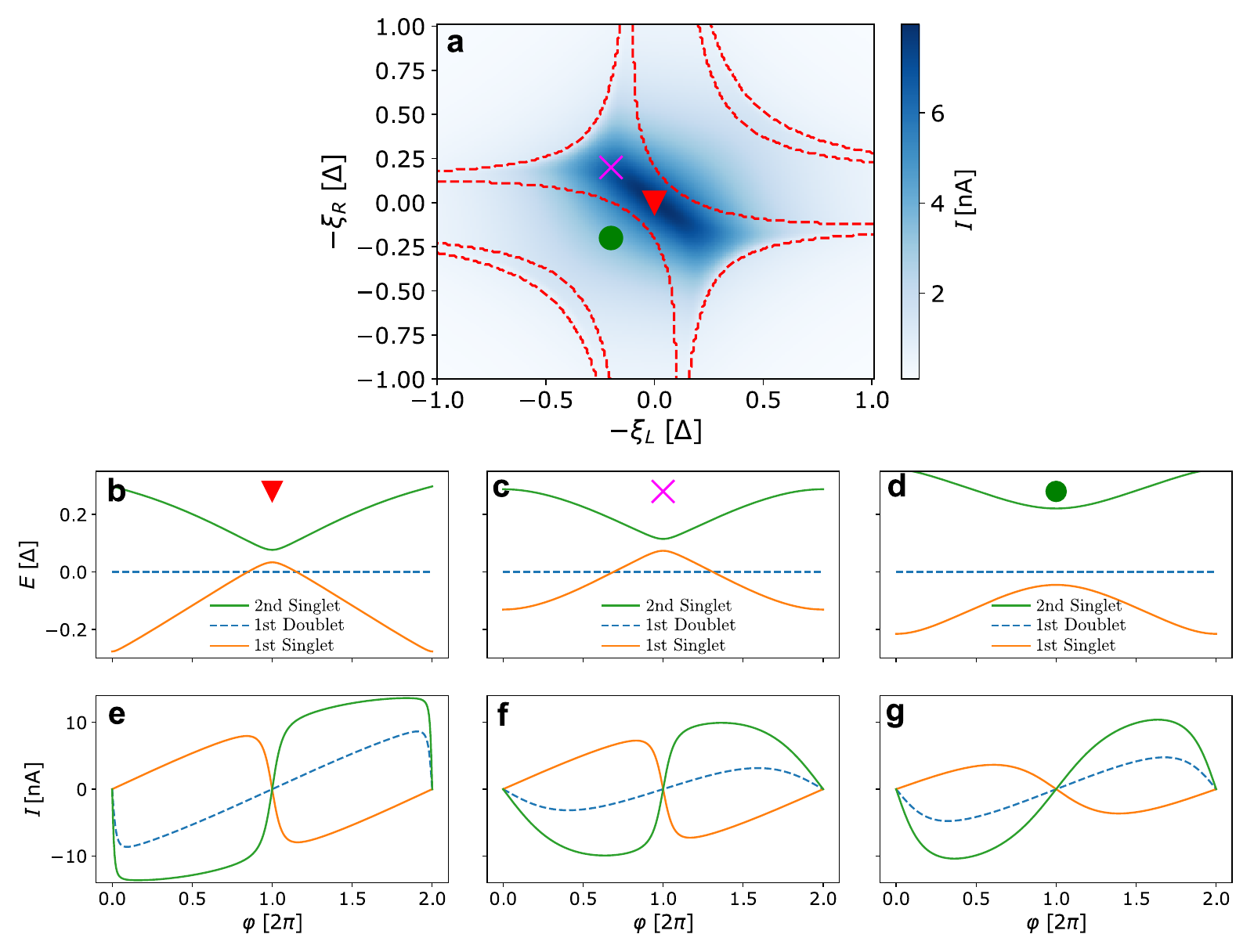}
    \caption{\textbf{a} Full range $I_c$ plot identical to Fig.~\ref{fig:6} of the main text. Icons indicate $\xi_L$ and $\xi_R$ settings in plots below. \textbf{b}-\textbf{d} State dispersion as a function of $\phi$ of the three lowest energy states for; \textbf{b} $\xi_L = \xi_R = 0.0$, \textbf{c} $\xi_L = 0.2\Delta$, $\xi_R = -0.2\Delta$, and \textbf{d} $\xi_L = \xi_R = 0.2\Delta$. Full lines indicate singlet parity states, while dashed lines indicate doublet parity. \textbf{e}-\textbf{g} Associated CPR of the three states plotted above. Color indicate state matching state in plots above.}
    \label{fig:SupTheory2}
\end{figure}

Next, we discuss the strong coupling regime shown in Fig.~\ref{fig:6} of the main text, and the dissimilarity between measured switching current and critical current in proximity to $\xi_L \approx \xi_R \approx 0.0$. In this regime, the current phase relation (CPR) becomes largely none-sinusoidal with groundstate transitions occurring as a function of $\phi$, as can be seen in Fig.~\ref{fig:SupTheory2}b-g. This is distinct from other explored regimes for which the CPR is mostly sinusoidal. Here, groundstate transitions are apparent in b and c, while d stays singlet for all $\phi$ and shows a more sinusoidal CPR in g. The skewed CPRs shown in e and f are related to small anti-crossings between the two lowest singlet states as seen in b and c. All this together leaves some complication in how to interpret the measured switching current. In a Resistively and capacitively shunted junction (RCSJ) model, the critical current branch with $V=0.0$ corresponds to a particle at rest in a washboard potential at a finite phase given by $\phi = \arcsin\left(I/I_c\right)$ \cite{tinkham2004introduction}. The stability of this state relates to the possibility of escaping into a running state ($V \neq 0.0$). For our system, in e.g. Figs.~\ref{fig:SupTheory2}c,f, the washboard potential would be morphed by the skewed CPR, and the lower energy of the doublet state in parts of the CPR would allow for groundstate transitions, depending on the singlet to doublet relaxation rate compared to the rate of $\phi$ change, typically dictated by circuit impedance at GHz frequencies. A full treatment of this is beyond our scope, and we simply note a correspondence between the range of the where model and experiment match well, with the occurrence of a groundstate transition as a function of $\phi$ which the red line marks in both Fig.~\ref{fig:6} and Fig.~\ref{fig:SupTheory2}.

\subsection*{Fitting procedure}
\begin{figure}[ht!]
    \centering
    \includegraphics{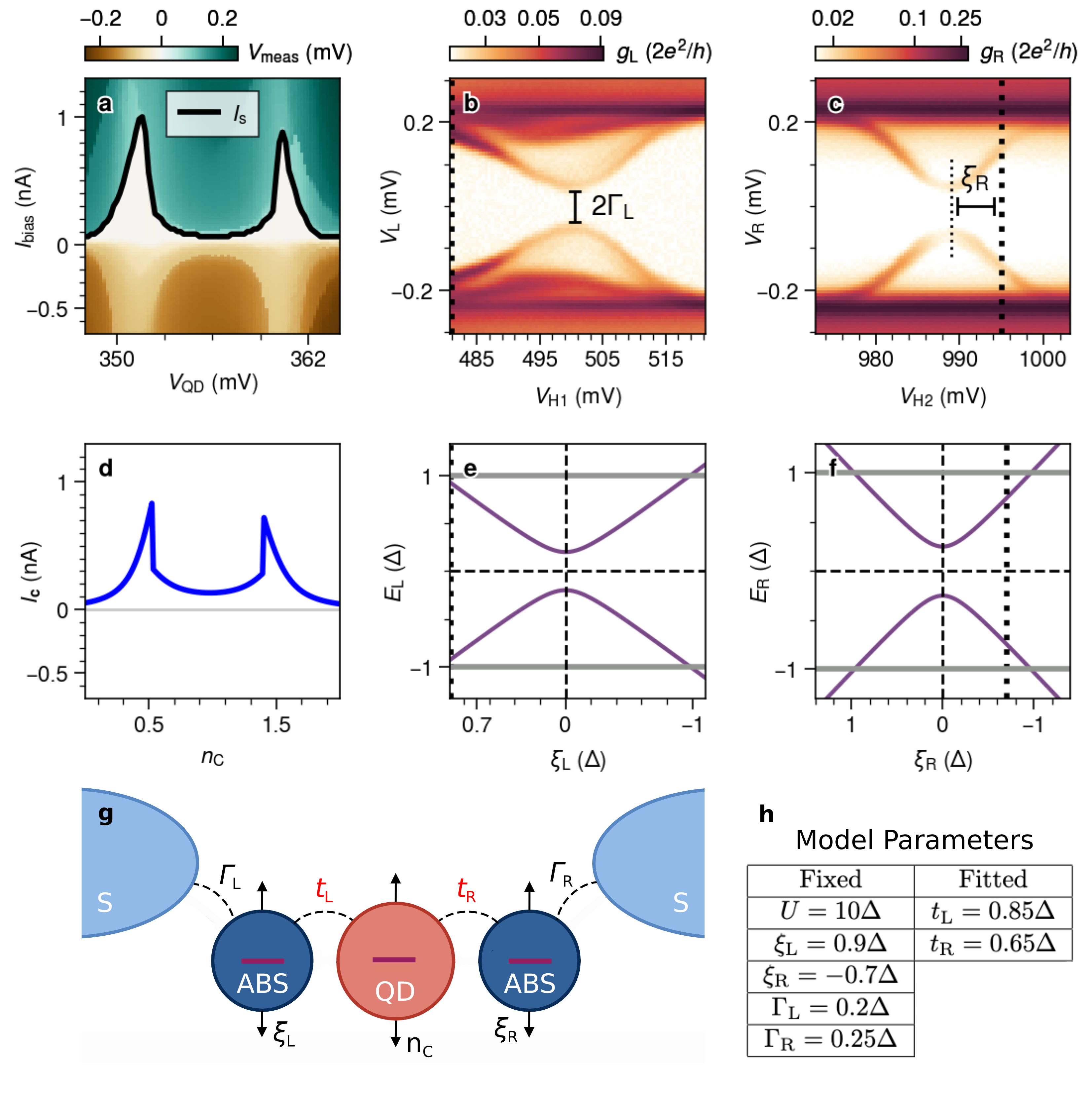}
    \caption{\textbf{The fitting procedure.}   \textbf{b.} $\Vmeas$ as a function of $\VQD$ and $\Ibias$. $\Vmeas$ is saturated at $\SI{250}{\micro V}$. The black line denotes $\Isw$, extracted at a threshold of $\SI{30}{\micro V}$. \textbf{b, c.}  The sub-gap excitation spectra measured from $N_{\mathrm{L}}$ and $N_{\mathrm{R}}$ as a function of $V_{\mathrm{H1}}$ and $V_{\mathrm{H2}}$ respectively. The dotted lines indicate the positions along $V_{\mathrm{H1}}$ and $V_{\mathrm{H2}}$ at which the supercurrent measurement of panel c was taken. \textbf{d.} A calculation of $I_{\mathrm{c}}$ as a function of $n_{\mathrm{C}}$, controlling the occupation of the QD. \textbf{e, f.} The excitation spectrum of the left and right ABS as a function of $\xi_{\mathrm{L}}$ and $\xi_{\mathrm{R}}$, simulated by the exact model. Grey horizontal lines indicate the gap edge. The dotted lines denote $\xi_{\mathrm{L}}$ and $\xi_{\mathrm{R}}$ as estimated from the experimental data in panel \textbf{b} and \textbf{c}.  \textbf{g.} A schematic of the model, depicting all parameters. \textbf{h.} The model parameters used for panel \textbf{d}. Fixed parameters are estimated based on ABS spectroscopy and Coulomb diamonds. }
    \label{supp:fitting-procedure}
\end{figure}
Estimating the theory model parameters relies on a variety of measurements, presented in Fig.~\ref{supp:fitting-procedure}a-c. The Hamiltonian parameters are illustrated in panel g. Both the ABS chemical potential $\xi_{\mathrm{L/R}}$ and the coupling between ABS and superconductor $\Gamma_{\mathrm{L/R}}$ can be estimated from ABS spectroscopy by comparing panels (b, e) and (c, f). We note that although the same ABSs are used throughout this work, $\Gamma_{\mathrm{L/R}}$ may vary slightly from figure to figure since it is sensitive to changes in the potential landscape generated by the gates, see details in the linked repository. Therefore, for every measurement presented in the main text, we monitor the ABS spectroscopy, from which we extract $\Gamma_{\mathrm{L/R}}$ and $\xi_{\mathrm{L/R}}$. The charging energy can be extracted from Coulomb diamonds presented in Fig.~\ref{fig:1}c, which leaves the couplings between the QD and the ABSs $t_{\mathrm{L/R}}$ as the only free parameters.  $t_{\mathrm{L/R}}$ are estimated based on the best fit to the experiment (panel a).

\subsection*{Rendering of 3D charge stability diagrams}

\begin{figure}[ht!]
    \centering
    \includegraphics[width=0.9\columnwidth]{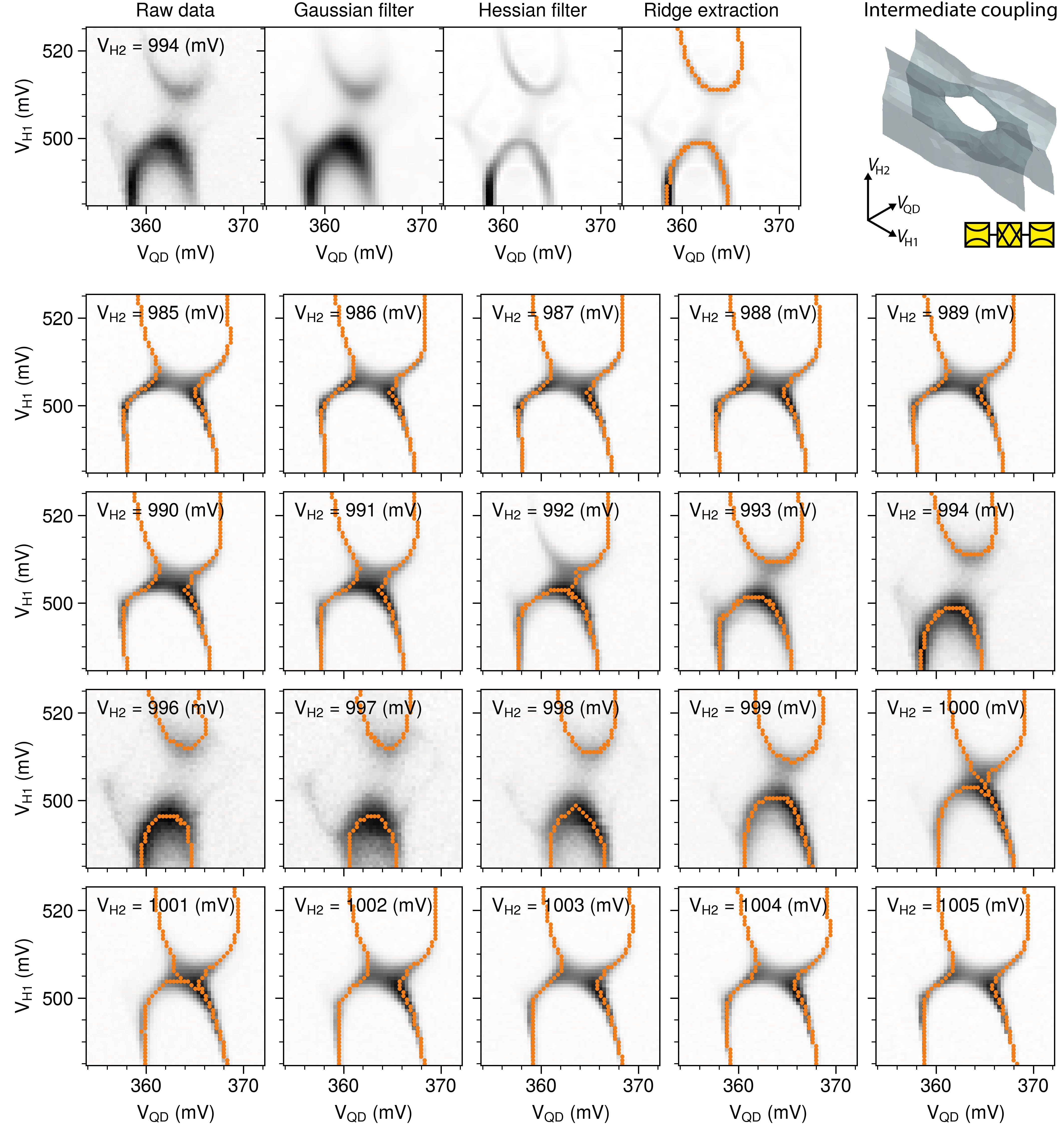}
    \caption{\textbf{3D charge stability diagram extraction procedure.} The four-panel sequence in the first row shows how the charge degeneracy points are extracted from a zero-bias conductance measurement. First, the raw data is smoothed with a Gaussian filter; then, a Hessian filter highlights the ridges; finally, a custom algorithm extracts the charge degeneracy points starting from the maxima on the top and bottom edges and following the pixels with maximum values. Such process is repeated for the 20 slices at different $\VHii$ values shown below; the extracted charge degeneracy points are plotted here on top of the raw conductance data. Eventually, all the charge degeneracy points are converted into the 3D manifold shown in the top-right corner using the \texttt{reconstruct\_surface} function of the \texttt{pyvista} python package. For further details see the Methods and the linked repository.}
    \label{supp:3D-CSD}
\end{figure}

\begin{figure}[ht!]
    \centering
    \includegraphics[width=0.5\columnwidth]{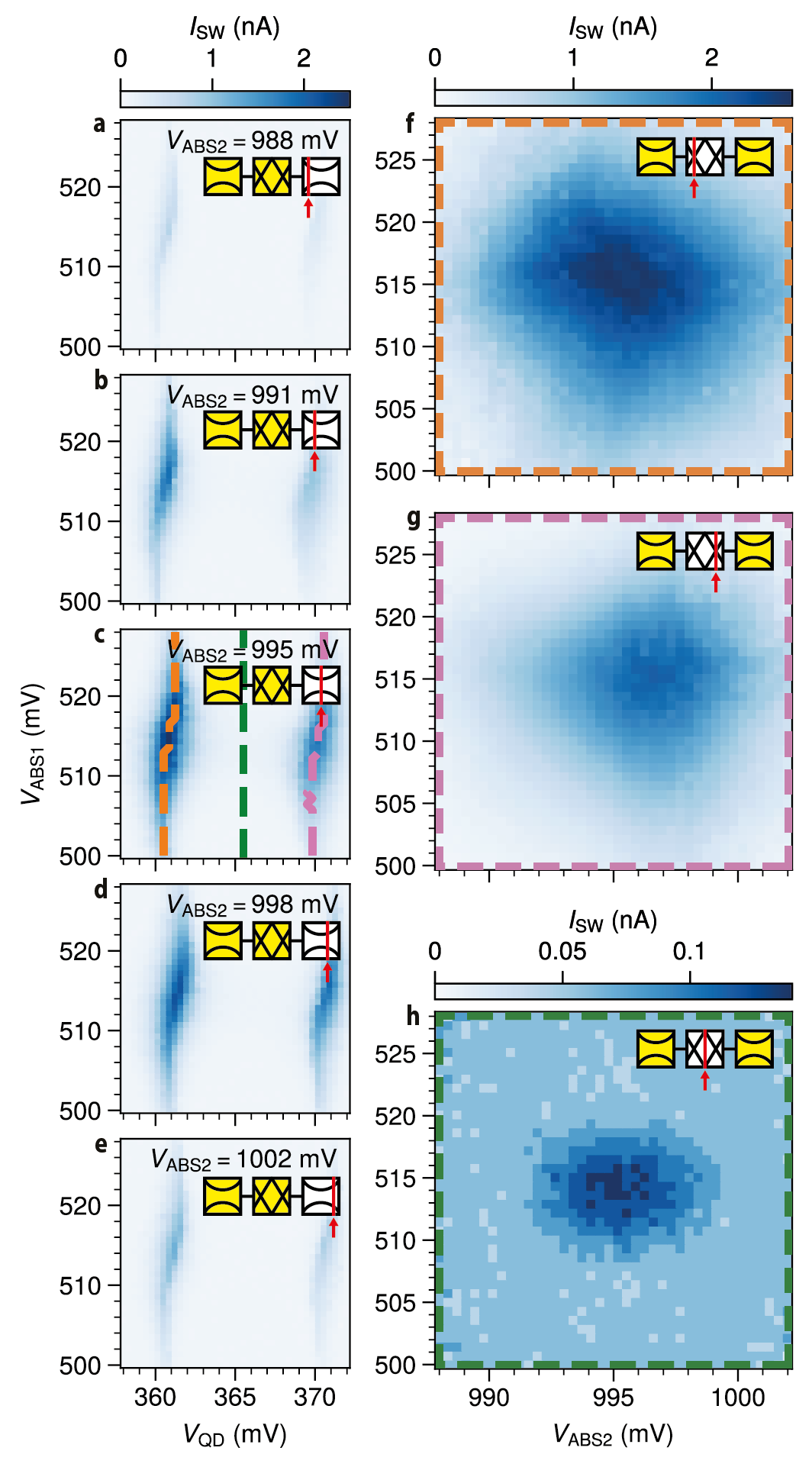}
    \caption{\textbf{a, b, c, d, e.} 2D maps of $\Isw$ as a function of $\VHi$ and $\VQD$. These are 5 examples among 40 $\VHi$-$\VQD$ maps taken at different $\VHii$ set points ranging from 988 to $\SI{1002}{mV}$, collectively forming a 3D dataset of $\Isw$ as a function of $\VHi$, $\VQD$ and $\VHii$. From such dataset we extract the data plotted in panels \textbf{f}, \textbf{g} and \textbf{h}. \textbf{f, g, h.} 2D maps of $\Isw$ as a function of $\VHi$ and $\VHii$. In panels \textbf{f} and \textbf{g}, $\VQD$ is placed along the left and right QD resonance respectively, colour-coded to superimposed lines in panel \textbf{c}. In panel \textbf{h}, $\VQD$ is placed in between the two QD resonances. Panel \textbf{f} reports the same data of Fig.~\ref{fig:3}g.}
    \label{supp:4D}
\end{figure}

Every 3D charge stability diagram of Fig.~\ref{fig:5} is extracted from a series of twenty $\VHi$–$\VQD$ 2D zero-bias conductance maps at different $\VHii$ set-points, as shown if Fig.~\ref{supp:3D-CSD}. From every 2D conductance map, we extract the charge degeneracy points with the following algorithm. First, the 2D map is smoothed with a Gaussian filter, then, a Hessian filter highlights the ridges by extracting the minimum eigenvalues of the matrix of second derivatives; for both filters, we use functions of the \texttt{scikit-image} python package. Finally, a custom \texttt{find\_ridge} routine extracts the charge degeneracy points from the filtered 2D map starting from the maxima on the top and bottom edges and following the pixels with maximum values. A representative example of this procedure is shown in the top row of Fig.~\ref{supp:3D-CSD}. The bottom part of Fig.~\ref{supp:3D-CSD} shows the result of the charge degeneracy point extraction on top of all the raw 2D conductance maps of the intermediate coupling regime. Similar plots for the weak and strong regimes are shared in the linked repository.
After all charge degeneracy points are extracted, they can be converted into a 3D manifold with a point-cloud-to-mesh conversion function. Specifically, we used the \texttt{reconstruct\_surface} function of the \texttt{pyvista} python package. 

All raw data, code and extracted charge degeneracy points for all coupling regimes are shared in the linked Zenodo repository.

\clearpage
\pagebreak
\newpage

\section*{Extended Data}

\begin{figure}[ht!]
    \centering
    \includegraphics{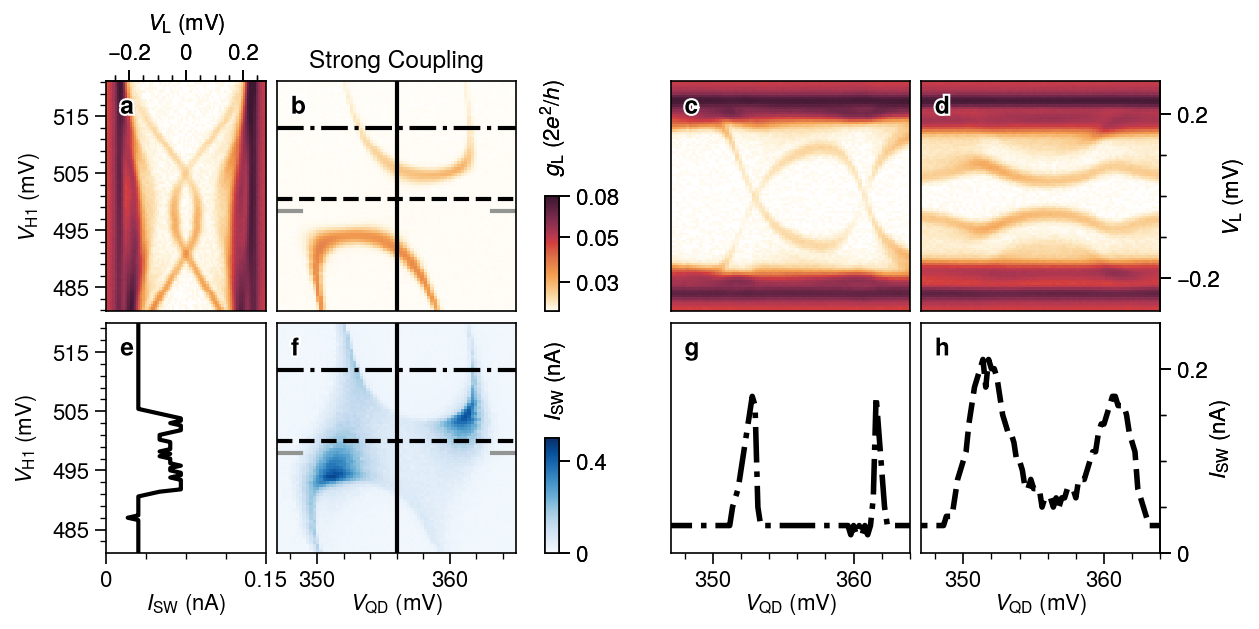}
    \caption{ \textbf{Dissection of the ABS-QD system in the strong coupling regime.} \textbf{a.} Spectroscopy as a function of $\VHi$ and measured from $\NL$. The position of $\VQD$ is indicated by the vertical line-cut in panel \textbf{b}. \textbf{b.} Zero-bias conductance, measured from $\NL$, as a function of $\VQD$ and $\VHi$. \textbf{c, d.} Spectroscopy as a function of $\VQD$ measured from $\NL$. Positions along $\VHi$ are indicated by horizontal lines in panel \textbf{b}. \textbf{e.} $\Isw$ as measured along the vertical black line of panel \textbf{f}. \textbf{f.} $\Isw$ as a function of $\VQD$ and $\VHi$. The gray inset indicates the minimum energy of the ABS. Dashed lines correspond to panels \textbf{g} and \textbf{h}. \textbf{g, h.} $\Isw$ as measured along the horizontal lines in panel \textbf{f}, extracted using a threshold of $\SI{30}{\micro V}$.}
    \label{supp:4D_StrongCoupling}
\end{figure}

\begin{figure}[ht!]
    \centering
    \includegraphics[scale=0.8]{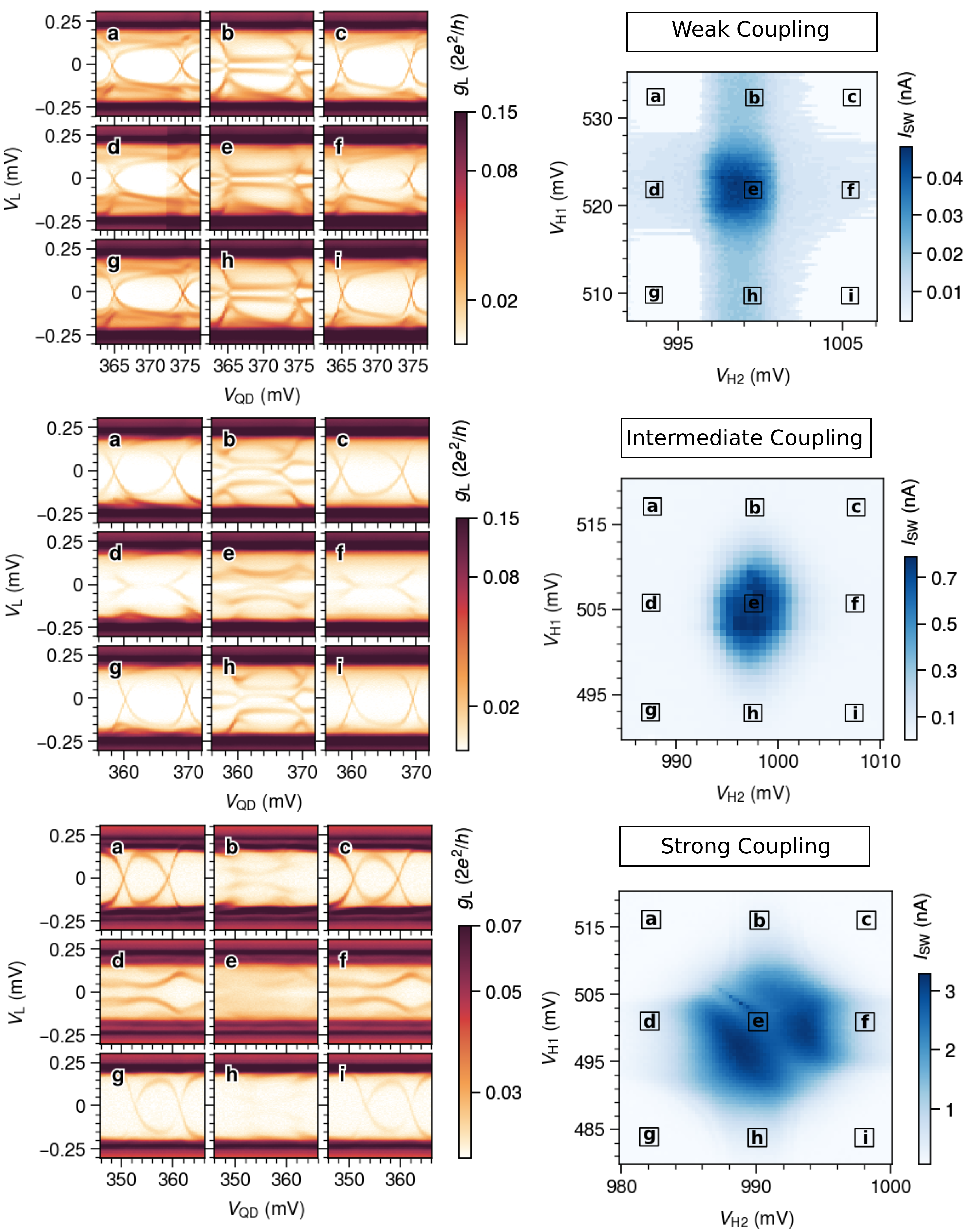}
    \caption{\textbf{Additional QD spectroscopy.} The right side of the figure shows three $\Isw$ maps as a function of both ABS gates in the weak, intermediate and strong coupling regime. The QD gate is kept between the two resonances as in Fig. \ref{fig:6}. The left side of the figure shows spectroscopy measured from $\NL$. Each plot corresponds to a different configuration of $\VHi$ and $\VHii$, as indicated in the $\Isw$ maps. In the weak coupling regime, parity switches are observed for every combination of $\VHi$ and $\VHii$. In the intermediate regime the parity switch no longer occurs when both ABSs are on resonance (panel e) resulting in a strong increase in the $\Isw$ map on the right. In the strong coupling regime, parity switches only occur when both ABSs are placed off resonance (panels a,c,g,i).}
    \label{supp:YSR-matrix}
\end{figure}

\begin{figure}[ht!]
    \centering
    \includegraphics[scale=0.8]{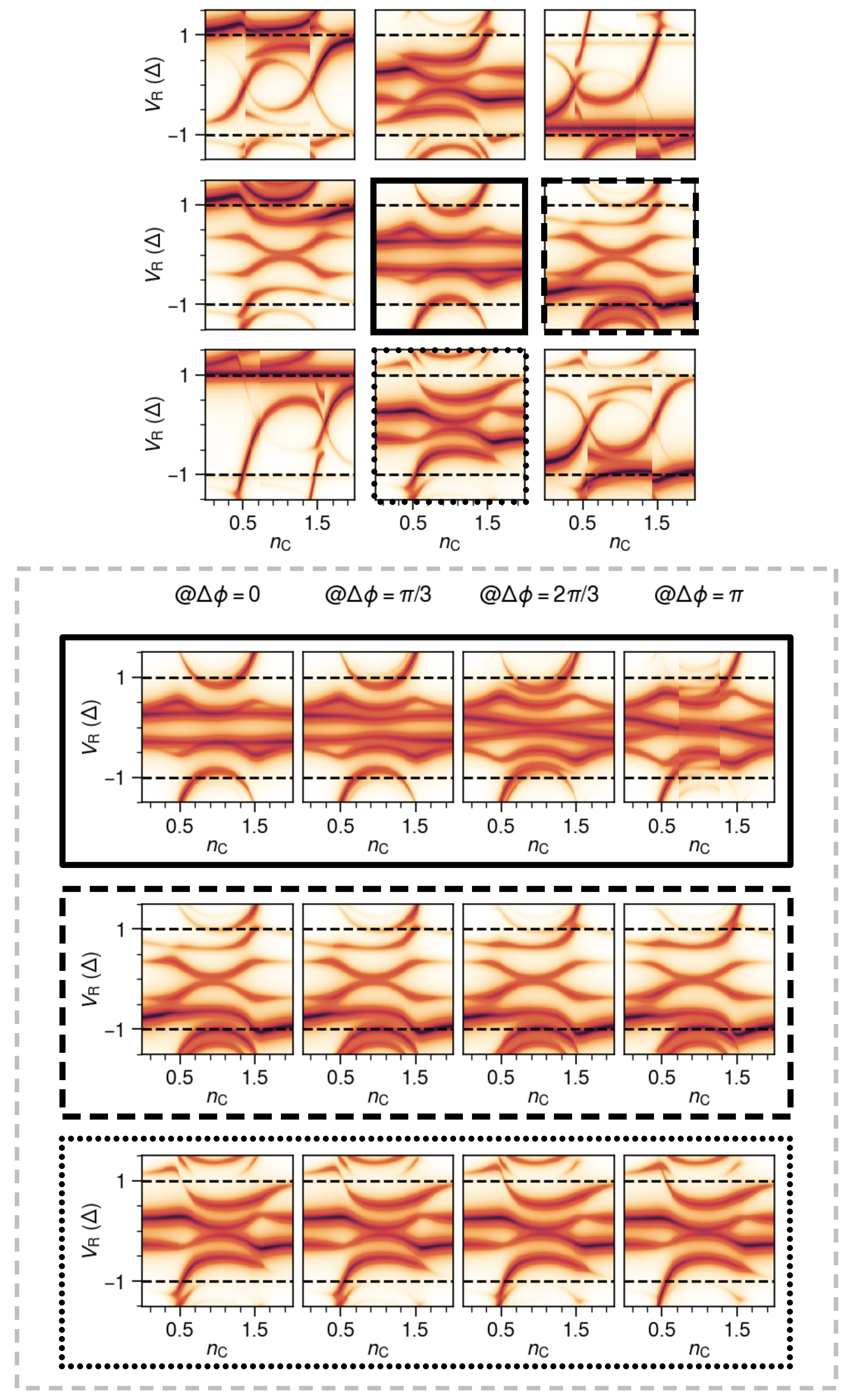}
    \caption{\textbf{Modulation of the excitation spectrum as a function of phase in the strong coupling regime.} Theory simulations of the zero-energy density of states, at $\phi = 0$, in accordance to the strong coupling data presented in Fig.~\ref{supp:YSR-matrix}. The middle panel, corresponding to both ABSs being on resonance and highlighted by a thick black line, is reevaluated at different values of $\phi$ in the lower half of the figure, ranging from $\phi=0$ to $\phi=\pi$. A modulation of the excitation spectrum is clearly visible in this case. Depending on $\phi$, the ground state parity of the system may be even ($\phi=0, \pi/3, 2\pi/3$) or odd ($\phi=\pi$). The region in $\VHi$ vs $\VHii$ space where this is possible is indicated in red in Fig. \ref{fig:6}.  The excitation spectrum of panels where one ABS is placed off resonance is largely unaffected by $\phi$, as can be observed from the other two panels presented in this figure.}
    \label{supp:phase}
\end{figure}

\begin{table}
    \centering
    \begin{tabular}{|c|c|c|c|l|l|l|l|l|} \hline 
         \textit{Fig.} &  $U$&  $n_{\mathrm{C}}$&  $\xi_{\mathrm{L}}$&$\xi_{\mathrm{R}}$ &$\Gamma_{\mathrm{L}}$ &$\Gamma_{\mathrm{R}}$ & $t_{\mathrm{L}}$&$t_{\mathrm{R}}$\\ \hline \hline
         Fig. 3d&  10&  [0.2, 1.8]&  -0.05, -0.35, -0.65& 0.37& 0.2& 0.16& 0.2&0.38\\ \hline 
 Fig. 3f& 10& -& [-0.8, 0.6]& 0.37& 0.2& 0.16& 0.18&0.46\\ \hline \hline
         Fig. 4g,j&  10&  [0,2]&  [-1,1]& 0.2& 0.3& 0.2& 0.18&0.15\\ \hline 
 Fig. 4h,k& 10& [0,2]& [-1,1]& 0.5& 0.25& 0.2& 0.6&0.27\\ \hline 
 Fig. 4i,l& 10& [0,2]& [-1,1]& 1.3& 0.2& 0.25& 0.7&0.45\\ \hline \hline
         Fig. 6&  10&  1&  [-1, 0.8]& [-1, 0.8]& 0.25& 0.35& 0.7&0.8\\ \hline \hline
         Fig. ED2a&  10&  [0,2]&  0& 0& 0.001& 0.001& 0.01&0.01\\ \hline 
 Fig. ED2b,c& 10& [0,2]& 0.2& 0& 1& 1& 0.4&0.1\\ \hline \hline
         Fig. ED3a&  10&  1&  [-1,1]& [-1,1]& 0.25& 0.35& 0.7&0.8\\ \hline 
 Fig. ED3b& -& -& 0& 0& -& -& -&-\\ \hline 
 Fig. ED3c& -& -& 0.2& -0.2& -& -& -&-\\ \hline 
 Fig. ED3d& -& -& 0.2& 0.2& -& -& -&-\\ \hline \hline 
         Fig. ED4&  10&  [0,2]&  0.9& -0.7& 0.2& 0.25& 0.85&0.65\\ \hline 
 Fig. ED9& 10& [0,2]& -& -& 0.3& 0.2& 0.8&0.75\\ \hline
    \end{tabular}
    \caption{A summary of all model parameters used throughout this text. All parameters except $n_C$ are in units of $\Delta = 0.2$~meV.}
    \label{tab:summary}
\end{table}

%TC:endignore

\end{document}